\documentclass[12pt]{article}
\usepackage{epsfig}
\usepackage{comment}
\usepackage{latexsym}
\usepackage{color}
\usepackage{amsmath}
\newcommand{\mysquare}[0]{\raise-.2ex\hbox{{\Large$\Box$}}}
\def\lsim{\mathrel{\rlap {\raise.5ex\hbox{$ < $}}
{\lower.5ex\hbox{$\sim$}}}}
\def\gsim{\mathrel{\rlap {\raise.5ex\hbox{$ > $}}
{\lower.5ex\hbox{$\sim$}}}} \topmargin -1.5cm \textheight=22.5cm \textwidth=16.5cm
\catcode`\@=11
\newcount\hour
\newcount\minute
\newtoks\amorpm
\hour=\time\divide\hour by60 \minute=\time{\multiply\hour by60
\global\advance\minute by-\hour}
\edef\standardtime{{\ifnum\hour<12 \global\amorpm={am}%
        \else\global\amorpm={pm}\advance\hour by-12 \fi
        \ifnum\hour=0 \hour=12 \fi
        \number\hour:\ifnum\minute<10 0\fi\number\minute\the\amorpm}}
\edef\militarytime{\number\hour:\ifnum\minute<10 0\fi\number\minute}
\def\draftlabel#1{{\@bsphack\if@filesw {\let\thepage\relax
   \xdef\@gtempa{\write\@auxout{\string
      \newlabel{#1}{{\@currentlabel}{\thepage}}}}}\@gtempa
   \if@nobreak \ifvmode\nobreak\fi\fi\fi\@esphack}
        \gdef\@eqnlabel{#1}}
\def\@eqnlabel{}
\def\@vacuum{}

\newcommand{\be}[0]{\begin{equation}}
\newcommand{\ee}[0]{\end{equation}}
\newcommand{\ba}[0]{\begin{eqnarray}}
\newcommand{\ea}[0]{\end{eqnarray}}

\def\bs{\begin{subequations}}
\def\es{\end{subequations}}

\def\thebibliography#1{%
\vskip 0.5cm \centerline{\bf \Large References}
\list{%
[\arabic{enumi}]}{\settowidth\labelwidth{[#1]} \leftmargin\labelwidth
\advance\leftmargin\labelsep
\usecounter{enumi}}
\def\newblock{\hskip .11em plus .33em minus .07em}
\sloppy\clubpenalty4000\widowpenalty4000 \sfcode`\.=1000\relax}

\renewcommand{\section}{\setcounter{equation}{0}\@startsection
{section}{1}{0mm}{-\baselineskip}{0.5\baselineskip} {\normalfont\Large\bfseries}}

\renewcommand{\subsection}{\@startsection
{subsection}{2}{0mm}{-\baselineskip}{0.5\baselineskip} {\normalfont\large\bfseries}}

\renewcommand{\subsubsection}{\@startsection
{subsubsection}{3}{0mm}{-\baselineskip}{0.5\baselineskip}
{\normalfont\normalsize\slshape}}

\usepackage{amssymb,amsfonts}
\usepackage{graphicx}
\usepackage{cite}

\def\bc{\begin{center}}
\def\ec{\end{center}}
\def\bea{\begin{eqnarray}}
\def\eea{\end{eqnarray}}

\def\F{{\cal F}}

\def\and{\quad\mbox{and}\quad}

\newcommand{\Z}{\mathbb{Z}}

\begin{document}
\begin{titlepage}
\begin{flushright}
LPTENS--08/65,
December
\end{flushright}

\vspace{2mm}

\begin{centering}
{\bf\huge Orbifold Symmetry Reductions of Massive Boson-Fermion Degeneracy\\
}
$~$\\

$~$\\
\vspace{.5 cm}
 {\Large \bf Ioannis Florakis and Costas Kounnas}

\vspace{2mm}

 {\Large  Laboratoire de Physique Th\'eorique,\\
Ecole Normale Sup\'erieure, \\
24 rue Lhomond, F--75231 Paris cedex 05, France\\}
{\em  Ioannis.Florakis@lpt.ens.fr\\ Costas.Kounnas@lpt.ens.fr}
\vspace {1.cm}

{\bf \Large Abstract}

\end{centering}
$~$\\
We investigate the existence of string vacua with {\it Massive Spectrum Degeneracy
Symmetry} ($MSDS$) in Heterotic and Type II orbifold constructions. We present a
classification of all possible $\Z_2^N$-orbifolds with $MSDS$ symmetry that can be constructed in the formalism of the 2d free
fermionic construction. We explicitly construct several two-dimensional models whose Reduced Massive Spectrum Degeneracy Symmetry ($RMSDS$) is due to a set of $\Z_2$-orbifold
projections induced naturally in the framework of the free fermionic construction. In all proposed models the massive  boson and fermion degrees of
freedom exhibit Massive Spectrum Degeneracy Symmetry while the number of massless
bosons $n(b)$ and massless fermions $n(f)$ are different; $n(b)\ne n(f)$.  This
property distinguishes the $MSDS$ $\Z_2$-twisted theories from ordinary supersymmetric
ones. Some comments are stated concerning the large marginal $J\bar
J$-deformations of the proposed models connecting them to higher-dimensional
gauged-supergravity theories with non-trivial geometrical fluxes.

\vspace{5pt} \vfill \hrule width 6.7cm \vskip.1mm{\small \small \small \noindent
Research
partially supported by the EU (under the contracts MRTN-CT-2004-005104,
MRTN-CT-2004-512194,
MRTN-CT-2004-503369,  CNRS PICS 2530, 3059 and ANR (CNRS-USAR) contract
 05-BLAN-0079-01. Unit{\'e} mixte  du CNRS et de l'Ecole Normale Sup{\'e}rieure
associ\'ee \`a
l'Universit\'e Pierre et Marie Curie (Paris
6), UMR 8549.}\\
\end{titlepage}
\newpage
\setcounter{footnote}{0}
\renewcommand{\thefootnote}{\arabic{footnote}}
 \setlength{\baselineskip}{.7cm}
\setcounter{section}{0}

\section{Introduction}
String theory provides a consistent framework that unifies all
interactions including gravity \cite{GSW}. Focusing ourselves on stringy
gravity and cosmology new interesting phenomena occur. Conventional
notions from general relativity like geometry, topology  etc.,
 are well defined in the string framework only as {\it low energy approximations} of
the stringy approach \cite{CosmoTopologyChange,GV,BV}.  At small distances
 physics deviates drastically from naive field-theoretic intuition. Various examples
of purely stringy phenomena have already been
 identified in the past, which in several cases imply that the physics at strong
curvature scales  can be quite different from what
 one might expect from the field theory approximation \cite{CosmoTopologyChange}.
 They indicate new
possibilities in the context of quantum cosmology and especially
 in the context of the ``Stringy  Big-Bang" picture \cite{CosmoTopologyChange,GV,BV}
 versus the initial time
singularity picture  of the ``Big-Bang" in General Relativity.
 Assuming for instance a compact space and sufficiently close to the singularity,
the typical scale of the universe reaches at these early
 times  the gravitational scale (string scale). Obviously at this early epoch
classical gravity is no longer valid and has to be replaced by
 a more fundamental singularity-free theory such as (super-)string theory.

 $~$\\
Changing our framework from field theory to strings is by far a non-straightforward
task, since even (super-)string theories are marked
by  Hagedorn-like singularities \cite{Hagedorn,AtickWitten,RostKounnas} which have to be resolved either by stringy phase
transitions \cite{AtickWitten,AKADK} or by choosing Hagedorn-free string vacua in the
early stage of the universe \cite{MSDS,GravFluxes}. From our viewpoint, it is of fundamental importance to
show that the space of Hagedorn-free vacua is not empty and
that their existence is {\it at least equally natural} as the Hagedorn-singular
ones. Recently, a  noticeable progress  has been made in constructing
Hagedorn-free string vacua in High Temperatures  in the presence of non-trivial
magnetic fluxes\cite{GravFluxes}, which has shown explicitly the existence of non-pathological
string vacua. Furthermore, a new symmetry ``Massive boson-fermion
Spectrum-Degeneracy Symmetry" ($MSDS$) was discovered in stringy vacua \cite{MSDS} where at
least eight of the nine space dimensions are compact with a typical compactification scale
close to the string scale.

 $~$\\
The fact that the $MSDS$ string vacua proposed in ref. \cite{{MSDS}} were constructed in
a $d\le 2$ target space background does not at all exclude
them from being the most serious candidates able to describe the early ``Stringy
non-geometric era'' of the universe. On the contrary, in the spirit of
ref. \cite{MSDS}, this is actually quite natural. Assuming compact transverse space,
the ``non-geometrical stringy era" is expected as a consequence of the
stringy $T$-duality symmetry. At the {\it$T$-self-dual points} the geometric
description of space breaks down. In stringy framework, however,  the theory makes
sense in terms of a ``non-geometrical description" based on non-abelian gauge field
theory. These stringy phenomena are well known in several stringy
compactifications  around the so called {\it``extended  gauge symmetry points"} of
the moduli space.

 $~$\\
 All $2d$ $MSDS$ string vacua, heterotic, type II and orientifolds proposed in ref.
\cite{MSDS} are non-geometrical in terms of  the internal compactified
 space but are well-characterized by the non-abelian  gauge group $H_L\times H_R$.
In the massless spectrum there are scalar
 bosons  $M_{I_L,J_R},~I_L=1,2,...,d_L,~J_R=1,2,...,d_R~$, parametrizing  the manifold
 \be
 \label{manifK}
 {\rm\cal K}={SO(d_L,~d_R)\over SO(d_L)\times SO(d_R)}\, ,
 \ee
where $ d_L,~d_R$ are the dimensions of the $H_L,~H_R$ gauge groups, respectively.
Because of the non-abelian structure of $H_L\times H_R$, the $MSDS$
vacua admit marginal deformations (flat directions) associated to the Cartan
sub-algebra $U(1)^{r_L}\times
U(1)^{r_R}$, with $r_L$ and $r_R$ being the ranks of  $H_L$ and  $H_R$, respectively.
Following ref. \cite{MSDS}, the moduli space of these deformations is of
current-current type $M_{IJ}~J^I_L\times J_R$ and is given by the coset :
 \be
 \label{manifM}
 {\rm\cal M}={SO(r_L,~r_R)\over SO(r_L) \times SO(r_R)}~.
 \ee
What is of main importance is the ultimate connection of the $M_{IJ}$ deformation
parameters with the {\it ``induced effective higher-dimensional space
geometry" } in  the {\it large $M_{IJ}$-deformation limit} (i.e. when the
$MSDS$-vacua are strongly-deformed). In this limit one recovers the geometric
field theory description in terms of  an effective  ``higher-dimensional"
conventional  superstring theory in which the space-time supersymmetry is
{\it spontaneously broken}  by ``geometrical" and ``thermal" fluxes
\cite{MSDS,GravFluxes,Cosmo-RT-Shifted}. This
fundamental generic  property of the  deformed $MSDS$-vacua strongly  suggests that
they be considered as the most (semi-) realistic candidate vacua able to describe
the  ``early non-singular phase of our Universe", being free of any
initial ``general relativity-like'' or ``Hagedorn-like'' stringy singularities.

 $~$\\
The originally proposed $MSDS$-vacua \cite{MSDS} and in particular the ones with $H_L\equiv
SU(2)^8$, are far too symmetric to be phenomenologically viable.
Indeed, in the extreme large-$M$ deformation limit (decompactification
limit), the induced effective theory is that of {\it non-chiral} extended
gauge supergravities, implemented with a well-defined  set of geometrical fluxes.
However, from our cosmological viewpoint,  the strongly deformed $MSDS$-vacua
should consistently represent our late time universe and, thus, should contain a 
non-trivial net number of chiral families as well as a reduced gauge group unifying
in the most realistic possible manner the standard model interactions.

 $~$\\
The main aim of this work is to show the existence of less symmetric
$MSDS$-vacua which are eventually connected via large $M$-deformations to
phenomenologically acceptable four-dimensional vacua.
In our days there are several well-known procedures that may be utilized in order to
reduce symmetries of string vacua and at the same time ensure the presence of
chiral matter representations of the unified gauge group. Such well-established
procedures that create chiral $N=1$
superstring models with (spontaneously) broken supersymmetry include symmetric
orbifolds \cite{orbifolds} ($\equiv $
Calabi-Yau \cite{GSW}) compactification,  fermionic constructions
\cite{ABK,KLT}, covariant lattices \cite{LLS} and Gepner
constructions\cite{GEPN}, asymmetric orbifolds\cite{ASorbifolds} ($\equiv$
generalized CY with torsion \cite{CYtorsion}), or type II orientifold
compactifications \cite{Orientifolds} with or without geometrical \cite{GeoFluxes} \cite{OpenFluxes}
or non-geometrical \cite{Fluxes} fluxes. In this work we apply
the ({\it Asymmetric  Freely Acting}) orbifold construction to $MSDS$-vacua, so that
the {\it ``Strongly $M$-Deformed $MSDS$-vacua" } one would obtain in late
 cosmological times be phenomenologically acceptable.

  $~$\\
The paper is organized as follows. In Section 2 a brief review of the construction
of maximally symmetric $MSDS$-vacua is presented. These theories are non-singular
and are based on a Spectral-Flow Super-Conformal Symmetry on the world-sheet. The
space-time spectrum exhibits a Massive Spectrum Degeneracy Symmetry ($MSDS$) between
massive bosons and fermions. In Sections 3 and 4 we employ fermionic and orbifold
construction techniques in order to construct several less symmetric models in Type
II (Section 3) and Heterotic (Section 4) theories, that are still characterized by a
{\it Reduced  Massive Spectrum Degeneracy Symmetry} ($RMSDS$).  In Section 5 we
derive the necessary conditions that permit the construction of all possible $RMSDS$-vacua, by utilizing free
fermionic construction techniques. The reduced moduli space of the models is studied
in section 6; in the same section we discuss the large $M$-deformation limit of
$RMSDS$-vacua and their plausible connections to phenomenologically acceptable
models in late cosmological times. Section 7 is devoted to our conclusions.


\section{Review of the maximally symmetric  $MSDS$-vacua}
In the maximally  symmetric  $MSDS$-vacua  all nine, or at least eight- space
coordinates are compact and closed to the string scale \cite{MSDS}.
Furthermore, all compact space coordinates are expressed in terms of free  2d
world-sheet fermions rather than the conventional compact
bosonic coordinates \cite{ABK,KLT,ABKW}.
The advantage of this fermionization lies in the consistent separation of left- and
right-moving world-sheet degrees of freedom into terms of left-
and right-moving 2d fermions that permit easier manipulations of the left-right
asymmetric (and even
non-geometrical) constructions of vacua in string theory. In what follows we
restrict ourselves to the case of two flat target-space dimensions, leaving the
remaining eight dimensions compactified to the string scale.
 \subsection{Generalities}
 \subsubsection{Type II  degrees of freedom }
In the ``critical"  Type II theories the left- and right- moving
degrees of freedom are:
  $~$\\
$\bullet$ The light-cone degrees: $(\partial X^0,~\Psi^0)$, $(\partial X^L,~\Psi^L)$
  $~$\\
$\bullet$ The super-reparametrization ghosts: $(b,c)$, $(\beta,\gamma)$
  $~$\\
$\bullet$ The transverse super-coordinates: $(\partial X^I,~\Psi^I),~I=1,...8$
 $~$\\
 In the fermionic construction the transverse super-coordinates are replaced by a
set  of free fermions in the adjoint representation of a semi-simple gauge group
 $H$ (refs.\,\cite{ABKW, ABK}): $\{\chi^a\},~a=1,...n$, $~n={\rm dim } [\, H\,]=24$. The
simplest choice of  $H$ is:
 \be
 H=SU(2)^8~\equiv~SO(4)^4,
 \ee
 where  the transverse super-coordinates $(\partial X^I, ~\Psi^I)$ are replaced by
 $(y^I,w^I,\Psi_I)$ so that  for every $I=1,...,8$,  the coordinate currents
$i\partial X^I =y^Iw^I$ are expressed in terms of the $y^I,w^I$ 2d world-sheet
fermions. For every $I$,  $\{y^I,w^I,\Psi^I\}$ define the adjoint representation of
a $SU(2)_{k=2}$ current algebra.
The choice $H= SU(2)^8$ of the coordinate-fermionization is by no means unique \cite{ABKW,MSDS}. Other
non-trivial  choices of the coordinate-fermionization are also possible:
 \be
 H=SU(5),~~ H=SO(7)\times SU(2),~~H=G_2\times Sp(4),~~H=SU(4)\times SU(2)^3,~~
H=SU(3)^3.
 \ee
 In this work we restrict our attention to $SU(2)^8$-fermionization  for both  left-
and right- moving transverse degrees of freedom.
 \subsubsection{Heterotic degrees of freedom }
The left-moving sector is identical to that of Type II theories, whereas the
right-moving degrees of freedom are \cite{MSDS}:  $~$\\
$\bullet$ The light-cone degrees: $(\partial X_0,~\partial X_L)$
  $~$\\
$\bullet$ The reparametrization ghosts: $(b,c) $
  $~$\\
$\bullet$ The transverse coordinates: $(\partial X^I,~I=1,...8)$
 $~$\\
 $\bullet$ The extra 32 right-moving fermions $(\psi^A,~A=1,2,...32)$ necessary for
the conformal anomaly cancelation. \\
In total there are 48 free fermions in the right moving sector $\{\bar
\chi^a,~a=1,2,...48\}$ which can be separated into :
 i) the 16 right-moving fermions $(y^I,w^I,~I=1,2,...,8)$  from coordinate
fermionization $i\partial X^I =y^Iw^I$ plus ii) the extra 32 right-moving fermions
$(\psi^A,~A=1,2,...32)$.
 \subsubsection{ The basic left- and right-moving chiral operators and partition functions}
The fundamental operators are as usual the left- and right-moving energy-momentum
tensor $T_B$ with conformal dimension $h_B=2$ and the left- and right-moving
superconformal operator $T_F$ with $h_F=3/2$ in Type II, responsible for the local left- and right-moving ${\rm
\cal N}=(1,1)$ world-sheet
superconformal symmetry \cite{GSW}. In Heterotic theories only the left-moving $T_F$
exists, giving rise to an ${\rm \cal N}=(1,0)$
superconformal symmetry.  In both Type II and Heterotic theories the left-moving
$T_B$ and $T_F$ have the same form:
 $$
 T_B=-{1\over 2}(\partial X_0)^2 -{1\over 2}\Psi_0\partial \Psi_0+ {1\over
2}(\partial X_L)^2+{1\over 2}\Psi_L\partial \Psi_L~+~  \sum_{a=1}^{24}~{1\over
2}~\chi^a \partial  \chi^a
 $$
 \be
T_F=i\partial X_0\Psi_0~+~i\partial X_1\Psi_1~+~\sum\limits_{a,b,c}{f_{abc}~\chi^a  \chi^b\chi^c}~,
 \ee
where $f_{abc}$ are the structure constants of the group $H_L=SU(2)^8$ and $
\{\chi^a\}$ $(a=1,2,...,24)$
denote the 8 fermionized super-coordinates:
\be
 \{\chi^a\},~a=1,2,...,24~~\equiv~~ \{\Psi^I ,y^I,w^I\},~I=1,2,...,8 \, .
 \ee
 The heterotic right-moving $\bar{T}_B(\bar{z})$ becomes :
 \be
 \bar{T}_B=-{1\over 2}(\bar \partial X_0)^2 + {1\over
2}(\bar \partial X_L)^2+
\sum_{a=1}^{48}~{1\over
2}~\bar \chi^a \bar \partial  \bar \chi^a\, .
 \ee
Following the rules of the fermionic construction and respecting the $H_{L}\times
H_R=SU(2)^8\times SU(2)^8$ in type II or the $H_L\times H_R=SU(2)^8\times SO(48)$
in the heterotic, we can construct very special tachyon free vacua,  with
left--right holomorphic
factorization of the partition function\cite{MSDS}. If the choice of  boundary
conditions on the world-sheet respects the  global existence of
the $H_L\times H_R$ symmetry, the latter is promoted to a local gauge symmetry on
the target  space-time, both in Type II and the Heterotic
cases \cite{ABKW, ABK, MSDS}. The simplest constructions are those where
all left-moving fermions $\{ \chi^a,~a=1,2,...24\}$  are taken with the same
boundary conditions.
All right-moving ones $\{\bar \chi^a,~a=1,2,...n_R\} $ have the same boundary
conditions as well ($n_R=24$ in Type II and $n_R=48$ in Heterotic). Both Type II and
Heterotic  partition functions
appear in simple factorized forms. In terms of the $SO(2n)$ characters ($n=12$ or
$n=24$):
\be
V_{2n}={\theta_3^n-\theta_4^n \over 2\eta^n },~~~~~  O_{2n}={\theta_3^n+\theta_4^n
\over 2\eta^n},~~~~~ S_{2n}={\theta_2^n-\theta_1^n \over 2\eta^n}, ~~~~~
C_{2n}={\theta_2^n+\theta_1^n \over 2\eta^n}\, ,
\ee
the Type II and Heterotic partition functions are:
$$
  Z_{II}=\int_\F {d^2\tau\over ({\rm Im}\tau)^2}~~
  \left(V_{24}^{\phantom{a^S}}-~S_{24} ~\right) \left (~\overline
V_{24}^{\phantom{a^S}}-~\overline S_{24} ~\right),
 $$
 \be
  Z_{Het}=\int_\F {d^2\tau\over ({\rm Im}\tau)^2}~~
  \left(V_{24}^{\phantom{a^S}}-~S_{24} ~\right) \left (~\overline
O_{48}^{\phantom{a^S}}+~\overline C_{48} ~\right).
 \ee
 The above expression for $ Z_{II}$ remains the same for any choice of
left- and right-moving $H$-group $H_L,~H_R$, since the dimension of each is always
equal to 24.
In this respect, {$ Z_{II}$ is a unique tachyon-free partition function (modulo the
chirality of the left-
and right-spinors) {\it that respects the $H_L\times H_R$ gauge symmetry}.
The expression of the left-moving part in $ Z_{Het}$  remains the same as well. The
right-moving part,
however, depends  on  the choice of $H_R$ (i.e. $SO(48),~ E_8\times SO(32),~E_8^3$).

$~$\\
Both $ Z_{II}$ and $ Z_{Het}$ show a Massive Spectrum Degeneracy Symmetry.
This spectacular property reflects the relations between the characters of the $SO(24)$-affine
algebra \cite{theta12,MSDS}:
 \be
  V_{24}~-~S_{24}={\rm constant} = 24~.
 \ee
 This follows from the well-known Jacobi identities between theta functions:
 $$
  \theta_3^{4}-\theta_4^{4} -  \theta_2^{4}=0, ~~~~\theta_1^{4}=0, ~~~~ \theta_2
\theta_3 \theta_4=2\eta^3,
 $$
 that further imply the identity:
 \be
 \label{theta12}
 {\theta_3 ^{12} - \theta_4 ^{12} \over 2\eta
^{12}}-~{\theta_2 ^{12} - \theta_1
^{12} \over 2\eta ^{12}}=24.
 \ee
  $~$\\
The above identity shows that the spectrum of  massive bosons and massive
fermions is identical to all string mass levels!  This is similar to the analogous property
of supersymmetic theories. In the massless level, however, the situation is
radically different: although  there are 24 left-moving   bosonic degrees of freedom there
are no massless fermionic states.

$~$\\
 $\bullet$ In Type II  there are 24 right-moving bosonic
states as well, so in total there are $24\times 24$ scalar bosons at the  massless
level
transforming under the adjoint representations of $H_L$ and $H_R$ .

  $~$\\
 $\bullet$ The integrated partition function is thus  equal to $[d(H_L)\times
d(H_R)]\times {\rm \cal
I} $, where ${\rm \cal I}$ is the integral over the fundamental domain
 $$
 {\rm \cal I}=\int_\F {d^2\tau\over ({\rm Im}\tau)^2}={\pi^2 \over 3},~~~~~~~~~
Z_{II}={\pi^2 \over 3}~ d(H_L)\times d(H_R).
 $$
 $\bullet$
 In  the  Heterotic string  the left-moving sector gives constant contribution as in the
Type II case $(d(H_L)=24)$.  The right-moving  massive states are
expressed in terms of the unique holomorphic modular invariant function  $j(\tau)$:
 \be
 Z_{Het}=\int_\F {d^2\tau\over ({\rm Im}\tau)^2} ~d(H_L)\times \left\{
d(H_R)+[j(\bar \tau)-744]\right\} ={\pi^2 \over 3}~
 d(H_L)\times d(H_R).
 \ee
The final integrated expression of  $Z_{Het}$ is similar to  $Z_{II}$. Both are
proportional to the number of massless states of the models. This is because the
contribution of the anti-holomorphic  function $[j(\bar \tau)-744]$  vanishes when integrated
over the fundamental domain. Depending on the choice of $H_R$ in the Heterotic,
the number of the massless states can be: $d[SO(48)]=1128,~ d[E_8\times SO(32)]=d[E_8^3]=744$.

\subsection{Chiral superconformal algebra and spectral flow in  $MSDS$}
 The symmetry operators of the $MSDS$ vacuum are the usual holomorphic
(anti-holomorphic) operators $T_B,~T_F$ ($\bar T_B,~\bar T_F$ ) giving rise to the
standard $\rm {\cal N}=$ $(1,1)$ world-sheet  superconformal symmetry in type II
and the
$\rm {\cal N}=$ $(1,0)$ in the heterotic, realizing a left-moving (and right-moving
in type II)
 Operator Product Expansion \cite{GSW} (OPE) with $\hat c={2\over 3}c=8$.
The extra symmetry operators are the currents of conformal weight  $h_J=1$,
associated with the
$H_L$- and $H_R$-affine algebras:
 $J^a \equiv f_{bc}^a \, \chi^b\chi^c$ and ${\bar J}^a  \equiv {\bar f}_{bc}^a
\,{\bar \chi}^b
{\bar \chi}^c$.
Finally, there are two $SO(24)$ spin-field  operators with conformal weight ${3\over
2}$ and opposite chirality :
\be
C=Sp\{\chi^a\}_+ ~~~~~~~{\rm and}~~~~~~S=Sp\{\chi^a\}_-
\ee
Following ref.\,\cite{MSDS}, the existence of the chiral operator $C$, of conformal
weight $h_C={3\over 2}$, together with $T_B,T_F, J^a, \chi^a$, form a {\it new
chiral
superconformal algebra} implying the massive boson-fermion degeneracy of the
spectrum. In order to show this, one needs to utilize the fusion rules as they are
described by the OPE relations between $C$ and $S$:
$$
C(z)~C(w)~{\sim}~{1\over (z-w)} \left\{{ {\rm \bf 1}\over (z-w)^2}~+~ { \hat \chi
\hat \chi \over
(z-w)}~+~ \dots \right\},
$$
$$
S(z)~S(w)~{\sim}~{1\over (z-w)} \left\{{{\rm \bf 1}\over (z-w)^2}~+~ { \hat \chi
\hat \chi  \over
(z-w)}~+~ \dots \right\},
$$
\be
\label{CS}
C(z)~S(w)~{\sim}~{1\over (z-w)^{1\over
2}}~\left\{{\hat \chi \over (z-w)^2}~+~{\partial
\hat \chi+\hat \chi \hat\chi \hat\chi\over
(z-w)}~+~ \dots \right\},
\ee
where $\hat \chi$ is a shorthand notation for $\gamma^a\chi_a$, with $\gamma^a$
being the
$\gamma$-matrices of $SO(24)$ . The above OPE relations between $T_B, T_F,
C_{\alpha}, J^a, \chi_a $ define  {\it a
new chiral superconformal algebra}.  The
$C(z)S(w)$ OPE in Eq.\,(\ref{CS}) implies a boson-fermion Spectral Flow which
guaranties the massive boson-fermion degeneracy of the Vacuum. The operator
$O_{3/2}\equiv(~ \partial \hat \chi~+~\hat \chi~\hat \chi~ \hat \chi ~)$ which
appears in the rhs of Eq.\,(\ref{CS}) is used to define  a massive bosonic vertex
operator of conformal weight $h_{1}= 2$ in the $(-1)$ ghost  picture \cite{FrShen,GSW}:
 \be
 {\bf V_{(1)}}\equiv e^{-\Phi}~(~ \partial \hat \chi~+~\hat \chi~\hat \chi~ \hat \chi ~).
 \ee
\subsubsection{Spectral flow and  the $MSDS$ operator-relations }
As usual the vertex operators are dressed by the super-reparametrization ghosts
\cite{FrShen,GSW}:  the space-time boson vertices are  expressed either in the
$0$ or the $(-1)$ ghost picture. The space-time fermions are in the $(-{1\over2})$ or $(-{3\over2})$
pictures.
\be
{\bf V_{(0)}}~=~ e^{-\Phi}~\hat\chi,   ~~~~~{\bf S}~=~ e^{-{1\over 2}\Phi
-{1\over 2}iH_0}~S~~{\rm or}~~{\bf S}~=~ e^{-{3\over 2}\Phi +{1\over 2}
iH_0}~S~,
\ee
where the $H_0$ is the usual helicity field  defined via bosonization of $\Psi_0 $
and $\Psi_L$: $i\partial H_0 =\Psi_0 \Psi_L$. The conformal weight $h_{q}$ of the
operator \cite{FrShen, GSW}:
\be
e^{q\Phi} ~~\longrightarrow~~h_{q}=-\,{1\over 2}\,q\,(q+2)
\ee
is such  that  ${\bf V_{(0)}}, {\bf V_{(1)}}$ have conformal weight $h_0=1,h_1=2$, while
${\bf S}$ has weight $h_S=2$ in both the $(-{1\over2})$ and $(-{3\over2})$ pictures. The string spectrum of bosons starts from a massless sector which is
described by $\bf V_{(0)}$.
On the contrary, all space-time fermions are massive, starting from mass level 1.
 The flow of ${\bf V_{(0),(1)}}$-states to ${\bf S}$-states
is expressed by the action of a ``Spectral-Flow Operator"  ${\bf C}$ :
$$
{\bf C}~\equiv~ e^{{1\over2}(\Phi-iH_0)}~C~.
$$
Here, ${\bf C}$ is written in the ($+\frac{1}{2}$) ghost picture. It has conformal dimension
$h_{\bf C}=1$ and $(-1/2)$ helicity charge. Thus, generically,
 ${\bf C}$ acting on ``physical"  bosonic states produces fermionic states at the same string level and
vice-versa. Although the ${\bf C}$-action looks like a
space-time supersymmetry transformation, the actual  situation turns out to be drastically
different from that of supersymmetry. Indeed, the ${\bf C}$-action leaves
the massless bosonic states of the theory invariant, therefore the boson-to-fermion
mapping does not exist for the massless states. This statement is visualized in the
OPE:
 \be
 \label{CV}
 {\bf C}(z)~{\bf V_0}~\sim~ {\bf S} \, ,~~{\rm finite~as ~}
z\rightarrow  w.
 \ee
The absence of singular terms in ($z-w$) shows clearly that the massless states
are invariant under the ${\bf C}$-transformation. On the other hand,  ${\bf C}$
acts not-trivially on the massive states:
\be
\label{CB}
{\bf C}(z)~{\bf V_1}(w) ~\sim~{{\bf S} (w) \over (z-w) }~ +~{\rm finite~terms}.
\ee
The above equation shows that the {\it massive bosonic states} are mapped into the
fermionic ones.  To show the inverse map ${\bf C}(z) :~{\bf S}(w) \rightarrow
{\bf V_{(1)} }(w)$, it is necessary to perform  standard picture-changing
manipulations \cite{FrShen,GSW} and consider  {\bf S} in the $(-\frac{3}{2})$ picture so that  ${\bf V_{(1)}}$
will appear in its conventional ghost-picture:
\be
\label{CSV}
{\bf C}(z)~{\bf S}(w) ~\sim~{{\bf V_{(1)}} (w) \over (z-w) }~ +~{\rm finite~terms}
\ee
$~$\\
 Summarizing: \\
$\bullet$    $T_B, T_F, C_{3/2}$ and $(J^a,\chi^a)$ define via the OPEs  a new
super-conformal algebra. \\
$\bullet$  The closure of the algebra is  guarantied when $c=12$, so that $C_{3/2}$
is a  chirality   ``$+$"  spin-feld of $SO(24)$  with conformal weight $h_C=3/2$ .
\\
$\bullet$  The realization of the algebra divides  the  ``physical"  states in two
sectors :\\
 i) A massless sector invariant under  ${\bf C}$ spectral-flow
transformations.\\
 ii) Massive fermionic states ${\bf S}$ with  ``$-$"  chirality, which are in one-to-one
correspondence with the massive bosonic states~   ${\bf C}$ :  ${\bf V_{(1)}}
\leftrightarrow  {\bf S}$ $~\longrightarrow$ {\it massive supersymmetry}.


\section{Type II $MSDS$ Orbifold Vacua}
In this section we provide  explicit examples of reduced $MSDS$ vacua ($RMSDS$), by
introducing $\mathbb{Z}_2$-twists and shifts on the internal compactified
 coordinates $ i\partial X^I\equiv y^I w^I $ of Type II theories. Our
constructions utilize the conformal field theory techniques of the free fermionic
construction \cite{ABK}, that can be easily translated in the symmetric and
asymmetric $\mathbb{Z}_2$-orbifold language. The orbifold representation will be
especially useful for the study of deformed  $RMSDS$ vacua via  $J \times \bar J$ marginal
deformations (see Section 6).

 \subsection{$\Z_2$-orbifolds with $MSDS$ in Type II theories}
\subsubsection{$\Z_2$-twisted $MSDS$ in Type II}
The initial  $MSDS$ vacuum in Type II factorizes the world-sheet fermions into two
basis sets
$\{H_L,H_R\}$:\\
i) the left-moving set
$~~~~~\rightarrow~~~~~H_L= \{\chi^{1\ldots
8},y^{1\ldots 8},w^{1\ldots 8}\}$ ,\\
 ii) the right-moving set
$~~\rightarrow~~~~~H_R=\{ \bar \chi^{1\ldots
8},\bar y^{1\ldots 8},\bar w^{1\ldots 8} \}.$
$~$\\
By introducing the additional (breaking) set:
$~~ B_t=\{\chi^{5\ldots
8} y^{5\ldots 8}|\bar{\chi}^{5\ldots 8}\bar{y}^{5\ldots 8}\}$,\\
the four internal coordinates are $\Z_2$-twisted as follows:
\be   i\partial X^I= y^I w^I \to - i\partial X^I=-y^I w^I ,~~i\bar\partial X^I= y^I
w^I \to - i \bar\partial X^I=-\bar y^I \bar w^I ,~~I=5,6,7,8.
\ee
 The partition function of the $\Z_2$-twisted  model $\{H_L,H_R, {B_t}\}$ is:
\be\label{Z2typeII}
Z_{II}^{B_t}~\equiv
~\frac{1}{2^2}\sum\limits_{a,b,\bar{a},\bar{b}}\frac{1}{2}\sum\limits_{h,g}{~(-)^{a+b}~\frac{\theta[^a_b]^8~\theta[^{a+h}_{b+g}]^4}{\eta^{12}}~~(-)^{\bar{a}+\bar{b}}~\frac{\bar{\theta}[^{\bar{a}}_{\bar{b}}]^8~\bar{\theta}[^{\bar{a}+h}_{\bar{b}+g}]^4}{\bar{\eta}^{12}}}~.
\ee
The $\Z_2$-projection induced by the  $B_t$ set, $\Z^{B_t}_2$-reduces the initial
spectrum symmetry:
\be
 SO(24)_L\times SO(24)_R~~ \longrightarrow ~~
 \left[ SO(16)\times SO(8)\right]_L\times \left[ SO(16)\times SO(8)\right]_R \, ,
 \ee
 so that the spectrum is naturally expressed in terms of the   $\left[ SO(16)\times
SO(8)\right]_{L,R}$ characters:
 \be
  V_{16},~O_{16},~S_{16},~C_{16},~~V_{8},~O_{8},~S_{8},~C_{8},~~
  \overline V_{16},~\overline O_{16},~\overline S_{16},~\overline C_{16},~~\overline
V_{8},~\overline O_{8},~\overline S_{8},~\overline C_{8}\, .
   \ee
The $\Z^{B_t}_2$-action is non-trivial on:
 \be
\Z^{B_t}_2 : ~ \{V_{8},~S_{8},~\overline V_{8},~\overline S_{8}\}~ \longrightarrow ~
- \{V_{8},~S_{8},~\overline V_{8},~\overline S_{8}\}.
 \ee
 The $Z_{II}^{B_t}$ partition function is naturally organized
into four products of holomorphic times anti-holomorphic terms
$\sum A_i\times\overline{A_i}$ transforming under the  same irreducible
representation of the $\Z^{B_t}_2$-group. There are two terms arising from the
untwisted sector $h=0$ and two from the twisted one $h=1$:
\be
Z_{II}^{B_t}=~\frac{1}{2}\sum\limits_{g}{Z_{+}[^0_g]\overline{Z}_{+}[^0_g]}+\frac{1}{2}\sum\limits_{g}{Z_{-}[^0_g]\overline{Z}_{-}[^0_g]}+\frac{1}{2}\sum\limits_{g}{Z_{+}[^1_g]\overline{Z}_{+}[^1_g]}+\frac{1}{2}\sum\limits_{g}{Z_{-}[^1_g]\overline{Z}_{-}[^1_g]}.
\ee
The first two terms, coming from the untwisted sector, are:
\be
 (i)~{\rm Untwisted}~(+,+):  ~~
\frac{1}{2}\sum\limits_{g}{Z_{+}[^0_g]\bar{Z}_{+}[^0_g]}=
\left(V_{16}O_8-S_{16}C_8 \right)\times\overline{\left(V_{16}O_8-S_{16}C_8\right)}\, ,
\ee
\be
 (ii)~{\rm Untwisted}~(-,-):  ~~
\frac{1}{2}\sum\limits_{g}{Z_{-}[^0_g]\bar{Z}_{-}[^0_g]}=
\left(O_{16}V_8-C_{16}S_8\right)\times\overline{\left(O_{16}V_8-C_{16}S_8\right)}\, .
\ee
The last two terms come from the twisted sector:
\be
 (iii)~{\rm Twisted}~(+,+):  ~~~~
\frac{1}{2}\sum\limits_{g}{Z_{+}[^1_g]\bar{Z}_{+}[^1_g]}=
\left(V_{16}C_8-S_{16}O_8\right)\times\overline{\left(V_{16}C_8-S_{16}O_8\right)} \, ,
\ee
\be
 (iv)~{\rm Twisted}~(-,-):  ~~~~
\frac{1}{2}\sum\limits_{g}{Z_{-}[^1_g]\bar{Z}_{-}[^1_g]}=
\left(O_{16}S_8-C_{16}V_8\right)\times\overline{\left(O_{16}S_8-C_{16}V_8\right)}\,.
\ee
It is remarkable that the $\{H_L,H_R, B_t\}$-twisted vacuum enjoys a massive
boson-fermion degeneracy symmetry similar to the original $\{H_L,H_R\}$ Type II model.
What is even more remarkable is that the $MSDS$ properties (reduced by $\Z^{B_t}_2$)
are not only valid  sector-by-sector individually but also  for each holomorphic and
anti-holomorphic factor separately. The (anti-)holomorphic part of each contribution
turns out
to be constant:
\be
\label{eqfirst}
    V_{16}O_8 - S_{16}C_8=16\, ,
\ee
\be
    O_{16}V_8 - C_{16}S_8=8\, ,
\ee
\be
    V_{16}C_8 - S_{16}O_8=0\, ,
\ee
\be
\label{eqlast}
    O_{16}S_8 - C_{16}V_8=8\, .
\ee
The above holomorphic {\it twisted} $\theta^{12}$-identities can be easily proved by
using the
``$\theta^4$-abstrusa" and ``triple-product'' identities (\ref{theta12}) of Jacobi.

$~$\\
 Adding all four contributions we obtain the partition function of the orbifolded
model:
\be
 Z_{II}^{B_t} =
16\times\overline{16}+8\times\overline{8}+0\times\overline{0}+8\times\overline{8}
= 384~.
\ee
 The fact that  the holomorphic and anti-holomorphic parts of each sector are
separately equal to constants, as shown in Eqs.\,(\ref{eqfirst}) - (\ref{eqlast}), implies that
the $MSDS$-structure originates separately from the holomorphic and
anti-holomorphic part in each sector. This precise property hints at the existence
of a \emph{chiral world-sheet algebra}, whose spectral-flow is responsible for the massive
boson-fermion degeneracy symmetry of the spectrum. In fact, the algebra in
question is a $\Z_2^{B_t}$-truncation of the original chiral
superconformal algebra of the previous section. The  $\Z_2^{B_t}$-orbifold truncates
the original spectral-flow operator ${\bf{C}}_{24}$, dual to the
$C_{24}$-character of $SO(24)$, down to the $\Z_2^{B_t}$-invariant operator:
\be
\Z_2^{B_t} ~:~C_{24} = C_{16} C_{8} +S_{16} S_{8}~ \longrightarrow~ C_{16} C_{8}
\ee
\be
{\bf C}_{{B_t}}=e^{{1\over2}(\Phi-iH_0)}~{C}_{16} \,{C}_{8}.
\ee
The global existence of ${\bf C}_{B_t}$, along with the truncated
chiral algebra, are sufficient to guarantee massive supersymmetry of
the spectrum.
$~$\\
We close this subsection with a few comments on the massless
spectrum of the $\Z_2^{B_t}$-orbifold  considered above. Its massless
states are still purely bosonic as a result of the left-right
symmetry of the model. Specifically, there are:
\begin{itemize}
 \item $16\times \overline{16}$ states:~ $\{\chi_{-\frac{1}{2}}^{1\ldots 4}\oplus y_{-\frac{1}{2}}^{1\ldots 4}\oplus
w_{-\frac{1}{2}}^{1\ldots 8}\}|0\rangle_L\otimes\{\overline{\chi}_{-\frac{1}{2}}^{1\ldots 4}\oplus \overline{y}_{-\frac{1}{2}}^{1\ldots 4}\oplus\overline{w}_{-\frac{1}{2}}^{1\ldots 8}\}|0\rangle_R$~
from $V_{16}O_8\times \overline{V}_{16}\overline{O}_8$
 \item $8\times \overline{8}$ states:~
$\{\chi_{-\frac{1}{2}}^{5\ldots 8}\oplus y_{-\frac{1}{2}}^{5\ldots 8}\}|0\rangle_L\otimes\{\overline{\chi}_{-\frac{1}{2}}^{5\ldots 8}\oplus\overline{y}_{-\frac{1}{2}}^{5\ldots 8}\}|0\rangle_R$~
from $O_{16}V_8\times\overline{O}_{16}\overline{V}_8$
 \item $8\times \overline{8}$ states:~ $Sp\{\chi^{5\ldots 8} y^{5\ldots 8}\}_{-}|0\rangle_L\otimes
Sp\{\overline{\chi}^{5\ldots 8} \overline{y}^{5\ldots 8}\}_{-}|0\rangle_R$~ from
$O_{16}S_8\times\overline{O}_{16}\overline{S}_8$
\end{itemize}
We note that each of the (anti-)holomorphic contributions to the
massless states come with a positive sign, implying that they are bosonic.

\subsubsection{$\Z_2$-shifted $MSDS$ in Type II}
A different way of reducing the initial $MSDS$ symmetry is via  $\Z_2$-shifted
orbifolds, where
some (or all) of the compactified coordinates are half-shifted :
\be
\Z_2^{B_s}~: ~ X^I\rightarrow X^I+\pi R \, .
\ee
In the fermionic construction the realization of $\Z_2$-shifted orbifolds is obtained
by the insertion of additional basis sets $B_s$ containing the world-sheet fermions
$\{y^Iw^I,\bar y^I\bar w^I\}$, which are associated with the shifted compactified
coordinates $X^I$.
In this example we will consider the case where all
the compactified coordinates are shifted ($I=1,2, \dots,8$), so that:
\be
B_s=\{y^{1\ldots 8} w^{1\ldots 8}\,|\,\overline{y}^{1\ldots 8}\overline{w}^{1\ldots
8}\} .
\ee
The partition function of the  $\{H_L,H_R, B_t\}$-shifted vacuum becomes:
\be
 Z_{II}^{B_s}=\frac{1}{2}\sum\limits_{h,g}\frac{1}{2^2}\sum\limits_{a,b,\bar{a},\bar{b}}{(-)^{a+b}~\frac{\theta[^a_b]^4~\theta[^{a+h}_{b+g}]^8}{\eta^{12}}~(-)^{\bar{a}+\bar{b}}~\frac{\bar{\theta}[^{\bar{a}}_{\bar{b}}]^4~\bar{\theta}[^{\bar{a}+h}_{\bar{b}+g}]^8}{\bar{\eta}^{12}}} ~.
\ee
Shifting the summation variables:
\be
a\rightarrow a+h~,~~ b\rightarrow b+g ~~;~~
\bar{a}\rightarrow \bar{a}+h~,~~ \bar{b}\rightarrow\bar{b}+g~,
\ee
we see that $Z_{II}^{B_s}$ becomes identical to the partition function $Z_{II}^{B_t}$ of the twisted model
(\ref{Z2typeII}).
The shift of variables produces an extra
phase $(-)^{h+g}$ in the holomorphic sector, which is cancelled by
the corresponding extra phase of the anti-holomorphic part and the partition
functions of the $\Z_2$-twisted and $\Z_2$-shifted models are algebraically equal.
However, this similarity between twists and shifts is due to the fact that the
theory has been written in the very special ``fermionic point'' of the moduli space.
The fundamental difference between the two models will become apparent as soon as
the theory is deformed away from that point. Indeed, the twisted model
corresponds to a non-freely acting orbifold, whose deformation radii have two fixed
points each, whereas the shifted model is freely-acting and effectively corresponds
to a model whose radii are reduced by half.

$~$\\
As in the twisted case, contributions to the partition function are naturally organized
into four products of holomorphic times anti-holomorphic terms
$\sum A_i\times\overline{A_i}$, transforming under the  same irreducible
representation of the $\Z^{B_s}_2$-group. Two of them arise from the untwisted
sector $h=0$ and two from the twisted one $h=1$:
$$
 (i)~~{\rm Untwisted}~(+,+):  ~  \frac{1}{2} \sum\limits_{g}
{Z_{+}[^0_g]\bar{Z}_{+}[^0_g]}=
 \left(V_8~O_{16}-S_8 C_{16}\right)\times\overline{\left(V_8 O_{16}-S_8
C_{16}\right)}=8\times\overline{8}
$$
$$
 (ii)~{\rm Untwisted}~(-,-):  ~
\frac{1}{2}\sum\limits_{g}{Z_{-}[^0_g]\bar{Z}_{-}[^0_g]}=\left(O_8
V_{16}-C_8 S_{16}\right)\times\overline{\left(O_8 V_{16}-C_8
S_{16}\right)}=16\times\overline{16}
$$
$$
 (iii)~~~{\rm Twisted}~(+,+):  ~
\frac{1}{2}\sum\limits_{g}{Z_{+}[^1_g]\bar{Z}_{+}[^1_g]}=\left(V_8
C_{16}-S_8 O_{16}\right)\times\overline{\left(V_8 C_{16}-S_8
O_{16}\right)}=(-8)\times\overline{(-8)}
$$
\be
(iv)~~~{\rm Twisted}~(-,-):  ~
\frac{1}{2}\sum\limits_{g}{Z_{-}[^1_g]\bar{Z}_{-}[^1_g]}=\left(O_8
S_{16}-C_8 V_{16}\right)\times\overline{\left(O_8 S_{16}-C_8
V_{16}\right)}=0\times\overline 0
\ee
where, as usual, the bars denote complex conjugation. We note
that the chiral and anti-chiral contributions of the shifted
 $(+,+)$-sector $\sum\limits_g{Z_{+}[^1_g]\bar{Z}_{+}[^1_g]}$ now come with a
 negative sign each, indicating an abundance of fermions in both the holomorphic and
anti-holomorphic sides. However, due to the  left-right symmetry of the model, the
 tensor product of the left- and right-handed spin fields
 $Sp\{\chi^{1\ldots 8}\}_{-}\otimes Sp\{\overline{\chi}^{1\ldots 8}\}_{-}$
 from $(-S_8 O_{16})\times(-\overline{S}_8 \overline{O}_{16})$
 produce space-time bosons, which accounts for the overall positive sign
$(-8)\times\overline{(-8)}=+64$.
The overall number of massless bosonic states is the same as in the
twisted case:
\be
    Z_{II}^{B_s}=16\times\overline{16}+8\times\overline{8}+(-8)\times\overline{(-8)}=384~.
\ee
As in the twisted case, the spectral-flow operator responsible for the
$MSDS$-symmetry is the truncation of the original operator,
invariant under the $\Z_2^{B_s}$-shift:
\be
 {\bf C}_{{B_t}}=e^{{1\over2}(\Phi-iH_0)}~{C}_{16} \,{C}_{8}~.
\ee

\subsubsection{$\Z_2$ left-twisted and right-shifted $MSDS$ in Type II}
Another interesting possibility in Type II is to combine together
an holomorphic $\Z_2$-twist with an anti-holomorphic $\Z_2$-shift.
Because of the chiral nature of the spectral-flow degeneracy, the
resulting model will again have a reduced $MSDS$-structure, since the
(anti-) holomorphic side does. The result will be an
asymmetric model whose additional basis  set $B_{ts}$ acts differently on the left-
and right-moving side:
\be
B_{ts}=\{\chi^{5\ldots 8}y^{5\ldots 8}\, | \,\overline{y}^{1\ldots
8}\overline{w}^{1\ldots 8}\}.
\ee
Its partition function is:
\be
\label{Z2typeIItwistshift}
 Z_{II}^{B_{ts}}=\frac{1}{2}\sum\limits_{h,g}\frac{1}{2^2}\sum\limits_{a,b,\bar{a},\bar{b}}{(-)^{a+b+hg}~\frac{\theta[^a_b]^8~\theta[^{a+h}_{b+g}]^4}{\eta^{12}}~(-)^{\bar{a}+\bar{b}}~\frac{\bar{\theta}[^{\bar{a}}_{\bar{b}}]^4~\bar{\theta}[^{\bar{a}+h}_{\bar{b}+g}]^8}{\bar{\eta}^{12}}}~.
\ee
The various contributions to the partition function are
organized as usual:
$$
(i)~~{\rm Untwisted}~(+,+):  ~
\frac{1}{2}\sum\limits_{g}{Z_{+}[^0_g]\bar{Z}_{+}[^0_g]}=\left(V_{16}O_8-S_{16}C_8\right)\times\overline{\left(V_8
O_{16}-S_8 C_{16}\right)}=16\times \overline{8}
$$
$$
(ii)~{\rm Untwisted}~(-,-):  ~
\frac{1}{2}\sum\limits_{g}{Z_{-}[^0_g]\bar{Z}_{-}[^0_g]}=\left(O_{16}V_8-C_{16}S_8\right)\times\overline{\left(O_8
 V_{16}-C_8 S_{16}\right)}=8\times\overline{16}
$$
$$
(iii)~~{\rm Twisted}~(+,+):  ~
\frac{1}{2}\sum\limits_{g}{Z_{+}[^1_g]\bar{Z}_{+}[^1_g]}=\left(O_{16}S_8-C_{16}V_8\right)\times\overline{\left(V_8
C_{16}-S_8
 O_{16}\right)}=8\times\overline{(-8)}
$$
\be
(iv)~~{\rm Twisted}~(-,-):  ~
\frac{1}{2}\sum\limits_{g}{Z_{-}[^1_g]\bar{Z}_{-}[^1_g]}=\left(V_{16}C_8-S_{16}O_8\right)\times\overline{\left(O_8
S_{16}-C_8
 V_{16}\right)}=0\times \overline{0}
\ee
The chiral and anti-chiral algebras need to be truncated by the
$\Z_2$-twist and $\Z_2$-shift, respectively. Similarly, the
holomorphic spectral-flow operator ${\bf{C}}_{B_{ts}}$ will be
invariant under the twist, while its anti-holomorphic counterpart
${\bf \overline C}_{B_{ts}}$ under the shift:
$$
  {\bf C}_{B_{ts}}=e^{{1\over2}(\Phi-iH_0)}~{C}_{16} \,{C}_{8}~,
$$
\be
{\bf \overline C}_{B_{ts}}=e^{{1\over2}(\bar\Phi-i\bar H_0)}~{\overline C}_{8}\,
{\overline C}_{16} ~ .
\ee
The full partition function of the model is then:
\be
 Z_{II}^{B_{ts}}=16\times\overline{8}+8\times\overline{16}+8\times\overline{(-8)}+0\times\overline{0}=192
\,
\ee
and exhibits massive supersymmetry, as expected. Note that the spectrum now contains
{\it massless fermions} in the twisted sector, as a result of the left-right
asymmetry of the model. There are $64$ massless states $Sp\{\chi^{5\ldots 8} y^{5\ldots
8}\}_{-}\otimes Sp\{\overline{\chi}^{1\ldots 8}\}_{-}$
arising from the sector $(O_{16}S_{8})\overline{(-S_8 O_{16})}$. Massless fermions
will also appear in the heterotic extension of (\ref{Z2typeIItwistshift}) which we
consider in Section 4.

\subsection{$\Z_2\times \Z_2$-orbifolds with $MSDS$ in Type II theories }
To further reduce the massive boson-fermion degeneracy of the initial Type II
maximally symmetric $MSDS$-vacuum, we next consider examples of $\Z_2\times
\Z_2$-orbifolds that exhibit Reduced Massive Spectral boson-fermion Degeneracy
Symmetry ($RMSDS$).
Here, as in the previous cases, the massive boson-fermion degeneracy follows from
(anti-)holomorphic $\Z_2\times \Z_2$-{\it twisted} $\theta^{12}$-identities
induced by an (anti-)holomorphic spectral-flow operator, invariant under
$\Z_2\times \Z_2$.

\subsubsection{$T^6\,/\,\Z_2\times \Z_2$ $MSDS$ orbifold in Type II}
In the language of the free fermionic construction, the $T^6\,/\,\Z_2\times \Z_2 $
orbifold of the original $MSDS$-vacuum is constructed by introducing two additional
basis sets $B_t^1,B_t^2$:
$$
 B_t^1=\{\chi^{5,6,7,8},y^{5,6,7,8}\,|\,\overline{\chi}^{5,6,7,8},\overline{y}^{5,6,7,8}\}\,,
$$
\be
 B_t^2=\{\chi^{3,4,7,8},y^{3,4,7,8}\,|\,\overline{\chi}^{3,4,7,8},\overline{y}^{3,4,7,8}\}\,,
\ee
giving different boundary conditions to the fermions $y^I$ and $w^I$, which bosonize
into the six compact scalars $i\partial X^I$, $I=3,\ldots,8$. The modular
invariant partition function of the $\{H_L,H_R,B_t^1,B_t^2\}$-vacuum is:
$$
Z_{II}^{B_t^1,B_t^2}=\frac{1}{2^2}\sum\limits_{h_i,g_i}\frac{1}{2^2}\sum\limits_{a,b,\bar{a},\bar{b}}
{(-)^{a+b}\over \eta^{12}}\times {(-)^{\bar{a}+\bar{b}}\over \bar\eta^{12} }
$$
\be
\theta\left[^a_b\right]^6~\theta\left[^{a+h_2}_{b+g_2}\right]^2~\theta\left[^{a+h_1}_{b+g_1}\right]^2~\theta\left[^{a-h_1-h_2}_{b-g_1-g_2}\right]^2\times
~\bar\theta\left[^{\bar{a}}_{\bar{b}}\right]^6~\bar
\theta\left[^{\bar{a}+h_2}_{\bar{b}+g_2}\right]^2~\bar\theta\left[^{\bar{a}+h_1}_{\bar{b}+g_1}\right]^2~\bar\theta\left[^{\bar{a}-h_1-h_2}_{\bar{b}-g_1-g_2}\right]^2~.
\ee
The $\Z_2^{B_t^1}\times \Z_2^{B_t^2}$ reduce the initial spectrum symmetry:
\be
SO(24)_{L,R}~~ \longrightarrow ~~
 \left[ SO(12)\times SO(4)\times SO(4)\times SO(4)\right]_{L,R}~,
\ee
 so that the spectrum is naturally expressed in terms of the $\left[ SO(12)\times
SO(4)\times SO(4)\times SO(4)\right]_{L,R} $ characters.
Taking the group elements of  $\Z_2^{B_t^1}\times \Z_2^{B_t^2}$ to be $\{\bf
1,a,b,ab\}$, we may organize
the contributions to the partition function into terms transforming as the
irreducible representations:
$$
 \Z_2^{B_t^1}\times \Z_2^{B_t^2} ~:~~~  \{ \chi_{(+,+,+,+)},
~\chi_{(+,+,-,-)},~\chi_{(+,-,+,-)},~\chi_{(+,-,-,+)}\}
 $$
 \be
 \longrightarrow  ~~~
   \{  {\bf 1}\, \chi_{(+,+,+,+)},~{\bf a}\,\chi_{(+,+,-,-)},~{\bf
b}\,\chi_{(+,-,+,-)},~{\bf ab}\,\chi_{(+,-,-,+)} \}~.
\ee
The transformation properties under  $\Z_2^{B_t^1}\times \Z_2^{B_t^2}$ of the
characters of the remaining spectrum  symmetry,  $\left[ SO(12)\times SO(4)\times
SO(4)\times SO(4)\right]_{L,R} $,  are:
$$
\{ C_{12}, ~C_{4}, ~ C_{4},~ C_{4}\} ~\longrightarrow ~\{ {\bf 1}\,C_{12}, ~{\bf
1}\,C_{4}, ~ {\bf 1}\, C_{4}, ~{\bf 1}\,C_{4}\} \, ,
$$
$$
\{ O_{12},~ O_{4},~  O_{4}, ~O_{4}\} ~ \longrightarrow ~\{ {\bf 1}\,O_{12},~ {\bf
1}\,O_{4},~  {\bf 1}\,O_{4}, ~{\bf 1}\,O_{4}\} \, ,
$$
$$
\{ \,V_{12}, ~V_{4}, ~ V_{4}, ~V_{4}\} ~~ \longrightarrow ~\{ {\bf 1}\,V_{12}, ~{\bf
a}\,V_{4}, ~{\bf b}\,V_{4}, ~{\bf ab}\,V_{4}\} \, ,
$$
\be
\{\, S_{12}, ~S_{4},  ~S_{4}, ~S_{4}\} ~~ \longrightarrow ~\{ {\bf 1}\,S_{12},~ {\bf
a}\,S_{4},  ~{\bf b}\,S_{4}, ~{\bf ab}\,S_{4}\}.
\ee
 The $Z_{II}^{B^1_t,B^2_t}$ partition function is naturally organized
into  products of sixteen holomorphic times anti-holomorphic terms
$\sum A_i\times\overline{A_i}$ transforming under the  same irreducible
representation of the $\Z_2^{B_t^1}\times \Z_2^{B_t^2}$-group. Four of those come
from the untwisted sector
$(h_1,h_2)=(0,0)$ and four from each of the three twisted planes
$(h_1,h_2)=(1,0)$, $(h_1,h_2)=(0,1)$  and
$(h_1,h_2)=(1,1)$.

$~$\\
{\bf {\it $\alpha$) Untwisted sector}  $(h_1,h_2)=(0,0)$}:
$$
i)~~\left| A_{(+,+,+,+)}^{(0,0)}\right|^2=\left|V_{12}O_4O_4 O_4+O_{12}V_4 V_4
V_4-S_{12}C_4 C_4 C_4-C_{12}S_4 S_4 S_4\right|^2
=12\times\overline{12}~,
$$
$$
ii)~~\left| A_{(+,-,+,-)}^{(0,0)} \right|^2=\left|V_{12}O_4 V_4 V_4+O_{12}V_4O_4 O_4
-S_{12}C_4 S_4 S_4 -C_{12}S_4 C_4 C_4 \right|^2
=~~4\times\overline{4}~,
$$
$$
iii)~~\left| A_{(+,+,-,-)}^{(0,0)} \right|^2=\left|V_{12} V_4 O_4V_4+O_{12}O_4
V_4O_4 -S_{12}S_4 C_4 S_4 -C_{12} C_4 S_4C_4 \right|^2
=~~4\times\overline{4}~,
$$
\be
~~~~iv)~~\left| A_{(+,-,-,+)}^{(0,0)} \right|^2=\left|V_{12} V_4 V_4O_4+O_{12}O_4
O_4V_4-S_{12}S_4 S_4C_4  -C_{12} C_4 C_4 S_4\right|^2
=~~4\times\overline{4} ~.~~~
\ee
The last three terms are algebraically equal due to the permutation symmetry under
the interchange of the three $SO(4)$ factors.
Summing the above, we obtain {\it  the untwisted contribution} to the partition
function:
\be
  \left| A_{(+,+,+,+)}^{(0,0)}\right|^2 +\left| A_{(+,-,+,-)}^{(0,0)}
\right|^2+\left| A_{(+,+,-,-)}^{(0,0)} \right|^2+\left| A_{(+,-,-,+)}^{(0,0)}
\right|^2= 12\times\overline{12}+(4\times\overline{4})\times3=192 ~.
\ee
The twisted contributions per twisted plane can again be grouped into four terms
according to their transformation properties.

$~$\\
{\bf {\it $\beta$) Twisted sector}  $(h_1,h_2)=(1,0)$}:
$$
i)~~\left| A_{(+,+,+,+)}^{(1,0)}\right|^2=\left| V_{12}O_4S_4 S_4+O_{12}V_4 C_4 C_4-
S_{12} C_4 V_4 V_4-C_{12} S_4O_4 O_4 \right|^2=~~0\times\overline{0}~,
$$
$$
ii)~~\left| A_{(+,+,-,-)}^{(1,0)}\right|^2=\left| V_{12}O_4 C_4 C_4+O_{12}V_4 S_4 S_4-
S_{12}C_4 O_4 O_4 -C_{12}S_4 V_4 V_4  \right|^2 =~~0\times\overline{0}~,
$$
$$
iii)~~\left| A_{(+,-,+,-)}^{(1,0)} \right|^2=\left| V_{12}
V_4 S_4 C_4+O_{12}O_4  C_4 S_4-S_{12}S_4V_4 O_4 -C_{12}C_4O_4 V_4  \right|^2
=~~4\times\overline{4}~,
$$
\be
~~~~iv)~~\left| A_{(+,-,-,+)}^{(1,0)} \right|^2=\left|V_{12}
V_4 C_4S_4 +O_{12}O_4 S_4 C_4-S_{12}O_4 V_4 S_4-C_{12}C_4V_4O_4  \right|^2
=~~4\times\overline{4} ~.~~~
\ee
{\bf {\it $\gamma$) Twisted sector}  $(h_1,h_2)=(0,1)$}:\\
The contribution of the $(h_1,h_2)=(0,1)$ twisted sector is similar to the
$(h_1,h_2)=(1,0)$
one. It is obtained by interchanging the characters of the first $SO(4)$ with the
second $SO(4)$ factor.

$~$\\
{\bf {\it $\delta$) Twisted sector}  $(h_1,h_2)=(1,1)$}:\\
The contribution of the $(h_1,h_2)=(1,1)$ twisted sector is similar to the
$(h_1,h_2)=(1,0)$ one.
 It is obtained by interchanging the characters of the first $SO(4)$ with the third
$SO(4)$ factor.
Thus, all three orbifold planes are equivalent, giving rise to equal contributions
per twisted sector, ($2\times (4\times 4)=32$). Therefore, the total partition
function of the  $\{H_L,H_R,B_t^1,B_t^2\}$-vacuum is constant:
\be
 Z_{II}^{B^1_t,B^2_t}~=~192+32\times 3=288~.
\ee
In the $\{H_L,H_R,B_t^1,B_t^2\}$-vacuum there are no massless fermions. The chiral
spectral-flow operator, invariant under $\Z_2^{B_t^1}\times \Z_2^{B_t^2}$, is:
\be
{\bf C}_{B^1_t , B^2_t}=e^{{1\over2}(\Phi-iH_0)}\,\left[\,{C}_{12} {C}_4
{C}_4 {C}_4 +{S}_{12}{S}_4 {S}_4 {S}_4\,\right]\, .
\ee

\subsubsection{$T^8$ /\,$\Z_2\times\Z_2$ $MSDS$ orbifold  in Type II}
We next discuss another example of $\Z_2\times\Z_2$ orbifold with a different
factorizable embedding in the compactified eight dimensional target space
$T^4/\Z_2\times T^4/\Z_2$.
Consider the orbifold constructed by the following choice of basis  sets:
$$
b^1_t=\{\chi^{5\ldots 8}y^{5\ldots 8}|\bar\chi^{5\ldots 8}\bar y^{5\ldots 8}\},
$$
\be
b^2_t=\{\chi^{1\ldots 4}y^{1\ldots 4}|\bar\chi^{1\ldots 4}\bar y^{1\ldots 4}\}.
\ee
The partition function of the $\{H_L,H_R,  b^1_t, b^2_t\}$-model is:
\be
 Z_{II}^{ b^1_t,
b^2_t}=\frac{1}{2^2}\sum\limits_{h_i,g_i}\frac{1}{2^2}\sum\limits_{a,b,\bar{a},\bar{b}}{(-)^{a+b}~
\frac{\theta{a\brack
b}^4~\theta{{a+h_2}\brack{b+g_2}}^4~\theta{{a+h_1}\brack{b+g_1}}^4}{\eta^{12}}}~(-)^{\bar{a}+\bar{b}}~
\frac{\bar\theta{{\bar{a}}\brack{\bar{b}}}^4~\bar\theta{{\bar{a}+h_2}\brack{\bar{b}+g_2}}^4~
\bar\theta{{\bar{a}+h_1}\brack{\bar{b}+g_1}}^4}{\bar\eta^{12}}~.
\ee
In this  model the initial spectrum symmetry  $SO(24)_L\times SO(24)_{R}$ is
reduced by $\Z^{ b^1_t}_2\times\Z^{ b^2_t}_2$ to a product of three $SO(8)$ factors:
\be
 SO(24)_{L,R}~~ \longrightarrow ~~
 \left[ SO(8)\times SO(8)\times SO(8)\right]_{L,R}~,
 \ee
so that the spectrum is naturally expressed in terms of the $\left[ SO(8)\times
SO(8)\times SO(8)\right]_{L,R} $ characters.
Taking the group elements of  $\Z_2^{b_t^1}\times \Z_2^{b_t^2}$ to be $\{\bf
1,a,b,ab\}$, we may organize
the contributions to the partition function into terms transforming as the
irreducible representations of the discrete orbifold group.
The transformation properties of the
characters of the remaining spectrum  symmetry  $\left[ SO(8)\times SO(8)\times
SO(8)\right]_{L,R} $ under  $\Z_2^{b_t^1}\times \Z_2^{b_t^2}$  are:
$$
\{ C_{8}, ~ C_{8},~ C_{8}\} ~\longrightarrow ~\{ {\bf 1}\,C_{8}, ~ {\bf 1}\, C_{8},
~{\bf 1}\,C_{8}\} \, ,
$$
$$
\{ O_{8},~  O_{8}, ~O_{8}\} ~ \longrightarrow ~\{ {\bf 1}\,O_{8},~  {\bf 1}\,O_{8},
~{\bf 1}\,O_{8}\} \, ,
$$
$$
\{ \,V_{8}, ~ V_{8}, ~V_{8}\} ~~ \longrightarrow ~\{ {\bf 1}\,V_{8}, ~{\bf
a}\,V_{8}, ~{\bf b}\,V_{8}\} \, ,
$$
\be
\{\, S_{8},  ~S_{8}, ~S_{8}\} ~~ \longrightarrow ~\{ {\bf 1}\,S_{8},  ~{\bf
a}\,S_{8}, ~{\bf b}\,S_{8}\}~.
\ee
 The $Z_{II}^{b^1_t,b^2_t}$ partition function is naturally organized
into  products of sixteen holomorphic times anti-holomorphic terms
$\sum A_i\times\overline{A_i}$ transforming under the  same irreducible
representation of the $\Z_2^{b_t^1}\times \Z_2^{b_t^2}$-group. Four of those come
from the untwisted sector
$(h_1,h_2)=(0,0)$ and four come from each of the three twisted planes
$(h_1,h_2)=(1,0)$, $(h_1,h_2)=(0,1)$  and
$(h_1,h_2)=(1,1)$.

$~$\\
{\bf \it $\alpha$) Untwisted sector $(h_1,h_2)=(0,0)$}:
$$
i)~~\left| A_{(+,+,+,+)}^{(0,0)}\right|^2=\left| V_8
O_8 O_8-S_8 C_8 C_8 \right|^2=8\times \bar{8}~,
$$
$$
ii)~~\left| A_{(+,+,-,-)}^{(0,0)}\right|^2 =\left| O_8
O_8 V_8-C_8 C_8 S_8\right|^2  =8\times \bar{8}~,
$$
$$
iii)~~\left| A_{(+,-,+,-)}^{(0,0)}\right|^2 =\left| O_8
V_8 O_8-C_8 S_8 C_8 \right|^2 =8\times \bar{8}~,
$$
\be
iv)~~\left| A_{(+,-,-,+)}^{(0,0)}\right|^2 =\left| \,V_8
V_8 V_8-S_8 S_8 S_8 \,\right|^2 =0\times \bar{0}~.
\ee
 $~$\\
{\bf \it $\beta$) Twisted sector $(h_1,h_2)=(1,0)$ }:
$$
i)~~\left| A_{(+,+,+,+)}^{(1,0)}\right|^2 =\left| V_8
C_8 O_8-S_8 O_8 C_8 \right|^2 =0\times \bar{0}~,
$$
$$
ii)~~\left| A_{(+,+,-,-)}^{(1,0)}\right|^2=\left| O_8
C_8 V_8-C_8 O_8 S_8  \right|^2=0\times \bar{0}~,
$$
$$
iii)~~\left| A_{(+,-,+,-)}^{(1,0)}\right|^2=\left| O_8
S_8 O_8-C_8 V_8 C_8  \right|^2=8\times \bar{8}~,
$$
\be
iv)~~\left| A_{(+,-,-,+)}^{(1,0)}\right|^2=\left| \, V_8
S_8 V_8-S_8 V_8 S_8 \, \right|^2=0\times \bar{0}~.
\label{usefuleq}
\ee
The two remaining twisted sectors are $(h_1,h_2)=(0,1)$ and $(h_1,h_2)=(1,1)$,
modulo permutations of the three $SO(8)$ factors. Summing up the contribution of the
untwisted
sector as well as the contribution of the three twisted sectors, the total partition
function of the $\{H_L,H_R,b_t^1,b_t^2\}$-vacuum  is:
\be
Z_{II}^{b^1_t,b^2_t}=192 + 64\times 3 = 384~.
\ee
The $\{H_L,H_R,B_t^1,B_t^2\}$-vacuum contains 384 massless bosons. There are no
massless fermions, as is the case in all examples of left-right symmetric orbifolds.
The chiral spectral-flow operator, invariant under $\Z_2^{b_t^1}\times \Z_2^{b_t^2}$
and responsible for the $MSDS$-structure, is:
\be
 { \bf{C}_{b^1_t,b^2_t}}=e^{{1\over2}(\Phi-iH_0)}\,C_8C_8C_8~.
\ee

\section{Heterotic  $MSDS$ Orbifold Vacua}
The fact that $MSDS$-structure is the result of a {\it chiral
spectral-flow} permits the construction of a large number of Type II
and Heterotic $MSDS$-vacua. It will be sufficient to
choose the holomorphic part of the partition function to be a
suitable twisting and/or shifting of the original model so that the
holomorphic contributions yield constants, while the
anti-holomorphic part is allowed to vary, respecting only the consistency conditions
of modular invariance. The resulting models will all
have $MSDS$-symmetry, since any non-constant contribution coming
from the anti-holomorphic side will necessarily violate
level-matching conditions and will, thus, not contribute to the
spectrum neither to the integrated partition function.

$~$\\
 In Heterotic $MSDS$-vacua the anti-holomorphic
contributions to the partition function are no longer constant
numbers and the full partition function becomes an anti-holomorphic
modular invariant function $Z_{het}(\bar{q}),~\bar q=\exp(-2i\pi\bar \tau)$.
It is well-known that any such function can be expressed as a rational function
$Q(\bar{j})$ of the Klein invariant $\bar{j}(\bar{\tau})$.
It is not difficult to see that the Klein $j$-function {\it can only
appear linearly} in the partition function as a result of the
structure of the anti-holomorphic part. Indeed, since the full partition function
will necessarily contain only simple would-be-tachyon poles $\sim 1/\bar{q}$ ,
the rational function $Q(\bar{j}(\bar\tau))$ is fixed to be at most
linear. The partition function will then necessarily be of the form:
\be
\label{genericHet}
 Z_{het}=n + m\left[\,\bar{j}(\bar\tau)-744\,\right]
\ee
 for some constant integers $n,m$.  It is straightforward  to show that $n$ equals
: $n(b)-n(f)$,
 namely the number of massless bosons minus the number of massless fermions of
the model, whereas $m$ is essentially the number of the would-be-tachyon poles in
$\bar{q}$. Moreover, it is clear that the integration over $\tau$ in the fundamental
domain
eliminates the spurious $[\bar{j}(\bar\tau)-744]$-terms leaving only the
constant contribution $n$ of the massless spectrum, as expected. In
what follows we give explicit examples of Heterotic models
with reduced  $MSDS$-structure, all of which are constructed in this spirit.

\subsection{Heterotic  $MSDS$ $\Z_2$-orbifolds }
The Heterotic $\Z_2$-orbifold models with $MSDS$-symmetry can be constructed by
coupling a holomorphic partition function $Z[^h_g]$ with $MSDS$-structure, such as
the ones studied in Section 3, to an
anti-holomorphic heterotic partition function $\bar{Z}[^h_g]$, so that the full
partition function be modular invariant:
\be
 Z_{het}=\frac{1}{2}\sum\limits_{h,g}{Z[^h_g]\bar Z[^h_g]}~.
\ee

\subsubsection{Heterotic $\Z_2$-twisted $MSDS$ with $SO(32)\times E_8$}
As a first example we consider the $\Z_2$-twisted holomorphic partition function:
$$
Z[^h_g]=\frac{1}{2}\sum\limits_{a,b}{(-)^{a+b+hg}~\frac{\theta[^a_b]^8\,\theta[^{a+h}_{b+g}]^4}{\eta^{12}}}~.
$$
For the right-moving side we choose an $SO(32)\times E_8$ gauge group:
$$
\bar{Z}[^h_g]=\frac{1}{2}\sum\limits_{\bar{a},\bar{b}}{\frac{\bar\theta[^{\bar{a}}_{\bar{b}}]^{16}\,
\bar\theta[^{\bar{a}+h}_{\bar{b}+g}]^8}{\bar\eta^{24}}
}~.
$$
This model is generated in the free fermionic construction by the following choice
of basis elements:
\be
H_L=\{\chi^{1\ldots 8},y^{1\ldots 8}, w^{1\ldots 8}\},~~~
H_R=\{\bar\chi^{1\ldots 48}\},~~~
B=\{\chi^{5\ldots 8}y^{5\ldots 8}|\bar\chi^{33\ldots 48}\}.
\ee
The contributions to the partition function are organized according to their
transformation under the $\Z^B_2$-twist, as usual:
$$
    \frac{1}{2}\sum\limits_{g}{Z_{(+,+)}[^0_g]\bar{Z}_{(+,+)}[^0_g]}=\left(V_{16}O_8-S_{16}C_8\right)\times\overline{\left(O_{32}O_{16}+C_{32}C_{16}\right)}=16\times
\overline{\left(O_{32}O_{16}+C_{32}C_{16}\right)}
$$
$$
    \frac{1}{2}\sum\limits_{g}{Z_{(+,-)}[^0_g]\bar{Z}_{(+,-)}[^0_g]}=\left(O_{16}V_8-C_{16}S_8\right)\times\overline{\left(V_{32}V_{16}+S_{32}S_{16}\right)}=8\times\overline{\left(V_{32}V_{16}+S_{32}S_{16}\right)}
$$
$$
    \frac{1}{2}\sum\limits_{g}{Z_{(+,+)}[^1_g]\bar{Z}_{(+,+)}[^1_g]}=\left(O_{16}S_8-C_{16}V_8\right)\times\overline{\left(O_{32}C_{16}+C_{32}O_{16}\right)}=8\times\overline{\left(O_{32}C_{16}+C_{32}O_{16}\right)}
 $$
\be
    \frac{1}{2}\sum\limits_{g}{Z_{(+,-)}[^1_g]\bar{Z}_{(+,-)}[^1_g]}=\left(V_{16}C_8-S_{16}O_8\right)\times\overline{\left(\,V_{32}S_{16}+S_{32}V_{16}\,\right)}=0\times\overline{\left(\,V_{32}S_{16}+S_{32}V_{16}\,\right)}
\ee
The holomorphic part for each individual term is constant due to the holomorphic
$MSDS$-structure, as explained in Section 3. To determine the full partition
function it is sufficient to derive the number of  massless
and would-be-tachyonic states. It is clear that massless states can only occur
from $V_{16}O_{8}\overline{O_{32}O_{16}}$, $O_{16}V_8\overline{V_{32}V_{16}}$ and
$O_{16}S_8\overline{O_{32}C_{16}}$. Specifically we have:
$$
    V_{16}O_{8}\overline{O_{32}O_{16}}~:16\times\overline{616}~~\textrm{states}~\{\chi_{-\frac{1}{2}}^{1\ldots
4}\oplus y_{-\frac{1}{2}}^{1\ldots 4}\oplus w_{-\frac{1}{2}}^{1\ldots
8}\}\otimes\{\bar\chi_{-\frac{1}{2}}^a\bar\chi_{-\frac{1}{2}}^b\oplus\bar\chi_{-\frac{1}{2}}^I\bar\chi_{-\frac{1}{2}}^J\}
$$
$$
    O_{16}V_8\overline{V_{32}V_{16}}~:~8\times\overline{512}~~\textrm{states}~\{\chi_{-\frac{1}{2}}^{5\ldots
8}\oplus y_{-\frac{1}{2}}^{5\ldots
8}\}\otimes\bar\chi_{-\frac{1}{2}}^a\bar\chi_{-\frac{1}{2}}^I
$$
$$
    O_{16}S_8\overline{O_{32}C_{16}}~:~8\times\overline{128}~~\textrm{states}~Sp\{\chi^{5\ldots
8}y^{5\ldots 8}\}_{-}\otimes Sp\{\bar\chi^I\}_{+}
$$
where $a,b=1\ldots 32$ span the fermions in $SO(32)$ while $I,J=33\ldots 48$ run
over the fermions in $SO(16)$.
There are in total $n=14976$ massless states in the model, while there is only one
anti-holomorphic tachyonic contribution
from $\overline{O_{32}O_{16}}$. After coupling to the holomorphic part, we obtain
the coefficient $m=16$ of the pole term.
Using (\ref{genericHet}) we determine the partition function of the model:
\be\label{heterZ2part}
    Z^B_{het}=14976+16\,\left[\,\bar{j}(\bar\tau)-744\right]~.
\ee
It is instructive to see how this structure appears at the level of characters.
Expanding the various contributions above in terms of $SO(8)$-characters and using
the triality relations $V_8=S_8=C_8$ we can write the partition function in a
particularly simple way:
$$
    Z^B_{het}=16\cdot\overline{\left[O_8^6+12 \,O_8^4 V_8^2+21 \,O_8^2 V_8^4 +
30\, V_8^6\right]}~.
$$
This expression should contain a $j$-function whose character expansion is:
\be\label{jcharacters}
    \bar{j}(\bar\tau) = (O_8^2+ V_8^2+S_8^2+C_8^2)^3=(O_8^2+3 V_8^2)^3~.
\ee
Using (\ref{jcharacters}) to eliminate the $O_8^6$-term in the partition function in
terms of the $j$-function, we find:
$$
    Z^B_{het} = 16\left[\bar{j}(\bar\tau)+3\,\overline{V_8^2(O_8^2-V_8^2)^2}\right]~.
$$
We next note that the second term is nothing but the square of $V_8 O_8 O_8 - S_8 C_8
C_8 = 8$
that was already encountered as (\ref{usefuleq}) in the $\Z_2\times\Z_2$ model with
triple triality. Therefore, the partition function becomes:
\be
    Z^B_{het} = 3072+
16\,\bar{j}(\bar\tau)=14976+16\,\left[\,\bar{j}(\bar\tau)-744\right]~,
\ee
which is, of course, the same as (\ref{heterZ2part}) found above by counting massless
and tachyonic states.

\subsubsection{Heterotic $\Z_2$-twisted $MSDS$ with $SO(16)\times SO(16)\times E_8$}
An interesting variation of the previous model comes from further breaking $SO(32)$
down to $SO(16)\times SO(16)$.
We consider the model generated by the following basis elements:
$$
    H_L=\{\chi^{1\ldots 8},y^{1\ldots 8}, w^{1\ldots 8}\}, ~~H_R=\{\bar\chi^{1\ldots
48}\},~~
    G=\{\bar\chi^{1\ldots 32}\},~~b=\{\chi^{5\ldots 8}y^{5\ldots
8}|\bar\chi^{16\ldots 32}\}.
$$
The partition function of the model is:
\be
    Z_{het}^{G,b}=
\frac{1}{2}\,\sum\limits_{h,g}\frac{1}{2}\,\sum\limits_{a,b}{(-)^{a+b+hg}~\frac{\theta[^a_b]^8\,\theta[^{a+h}_{b+g}]^4}{\eta^{12}}
}\cdot\frac{1}{2^2}\sum\limits_{\bar{a},\bar{b},\gamma,\delta}{\frac{\bar\theta[^{\bar{a}}_{\bar{b}}]^{8}\,\bar\theta[^{\bar{a}+h}_{\bar{b}+g}]^8\,\bar\theta[^\gamma_\delta]^8}{\bar\eta^{24}}
}~.
\ee
The partition function in terms of $SO(16)$-characters is organized into the
following contributions:
$$
    \frac{1}{2}\sum\limits_{g}{Z_{(+,+)}[^0_g]\bar{Z}_{(+,+)}[^0_g]}=\left(V_{16}O_8-S_{16}C_8\right)\times\overline{\left(O_{16}O_{16}O_{16}+O_{16}O_{16}C_{16}+C_{16}C_{16}O_{16}+C_{16}C_{16}C_{16}\right)}
$$
$$
    \frac{1}{2}\sum\limits_{g}{Z_{(+,-)}[^0_g]\bar{Z}_{(+,-)}[^0_g]}=\left(O_{16}V_8-C_{16}S_8\right)\times\overline{\left(V_{16}V_{16}O_{16}+V_{16}V_{16}C_{16}+S_{16}S_{16}O_{16}+S_{16}S_{16}C_{16}\right)}
$$
$$
    \frac{1}{2}\sum\limits_{g}{Z_{(+,+)}[^1_g]\bar{Z}_{(+,+)}[^1_g]}=\left(O_{16}S_8-C_{16}V_8\right)\times\overline{\left(O_{16}C_{16}O_{16}+O_{16}C_{16}C_{16}+C_{16}O_{16}O_{16}+C_{16}O_{16}C_{16}\right)}
$$
$$
    \frac{1}{2}\sum\limits_{g}{Z_{(+,-)}[^1_g]\bar{Z}_{(+,-)}[^1_g]}=\left(V_{16}C_8-S_{16}O_8\right)\times\overline{\left(V_{16}S_{16}O_{16}+V_{16}S_{16}C_{16}+S_{16}V_{16}O_{16}+S_{16}V_{16}C_{16}\right)}
$$
As before, the holomorphic part for each individual term is constant (and equal to 16, 8, 8 and 0, respectively)
due to the holomorphic $MSDS$-structure, as shown in Section 3. To determine
the full partition function it is sufficient to derive the number of  massless and
would be tachyonic states. Specifically the
massless spectrum is:
$$
    V_{16}O_{8}\overline{O_{16}O_{16}O_{16}}~:16\times\overline{360}~~\textrm{states}~\{\chi_{-\frac{1}{2}}^{1\ldots
4}\oplus y_{-\frac{1}{2}}^{1\ldots 4}\oplus w_{-\frac{1}{2}}^{1\ldots
8}\}\otimes\{\bar\chi_{-\frac{1}{2}}^a\bar\chi_{-\frac{1}{2}}^b\oplus\bar\chi_{-\frac{1}{2}}^I\bar\chi_{-\frac{1}{2}}^J\oplus\bar\chi_{-\frac{1}{2}}^\alpha\bar\chi_{-\frac{1}{2}}^\beta\}
$$
$$
    V_{16}O_{8}\overline{O_{16}O_{16}C_{16}}~:16\times\overline{128}~~\textrm{states}~\{\chi_{-\frac{1}{2}}^{1\ldots
4}\oplus y_{-\frac{1}{2}}^{1\ldots 4}\oplus w_{-\frac{1}{2}}^{1\ldots
8}\}\otimes Sp\{\bar\chi^\alpha\}_{+}
$$
$$
    O_{16}V_8\overline{V_{16}V_{16}O_{16}}~:~8\times\overline{256}~~\textrm{states}~\{\chi_{-\frac{1}{2}}^{5\ldots
8}\oplus y_{-\frac{1}{2}}^{5\ldots
8}\}\otimes\bar\chi_{-\frac{1}{2}}^a\bar\chi_{-\frac{1}{2}}^I
$$
$$
    O_{16}S_8\overline{O_{16}C_{16}O_{16}}~:~8\times\overline{128}~~\textrm{states}~Sp\{\chi^{5\ldots
8}y^{5\ldots 8}\}_{-}\otimes Sp\{\bar\chi^I\}_{+}
$$
$$
    O_{16}S_8\overline{C_{16}O_{16}O_{16}}~:~8\times\overline{128}~~\textrm{states}~Sp\{\chi^{5\ldots
8}y^{5\ldots 8}\}_{-}\otimes Sp\{\bar\chi^a\}_{+}
$$
The contribution of the massless states is therefore $n=11904$. Moreover, there are
$16$ would-be-tachyonic states from $V_{16}O_{8}\overline{O_{16}O_{16}O_{16}}$
giving a pole contribution $m=16$. Therefore, the partition function of the model is:
\be
    Z_{het}^{G,b} =
16\,\bar{j}(\bar\tau)=11904+16\,\left[\,\bar{j}(\bar\tau)-744\right]~.
\ee
We see that the heterotic $\Z_2^{G,b}$-twisted model has no massless fermions, since
those can only appear from the holomorphic side. The situation changes if one considers
a holomorphic $\Z_2$-shift, as shown below.

\subsubsection{Heterotic $\Z_2$-shifted $MSDS$ with $SO(32)\times E_8$}
We next illustrate an example of Heterotic $MSDS$-vacuum with massless fermions. We
consider the model that couples the $\Z_2$-shifted holomorphic $MSDS$-partition
function to the anti-holomorphic side with gauge group $SO(32)\times E_8$. The breaking set
of the shifted model is:
\be
B_s=\{y^{1\ldots 8}w^{1\ldots 8}\,|\,\bar\chi^{33\ldots 48}\}
\ee
and the partition function becomes:
\be
    Z^{B_s}_{het} =
\frac{1}{2}\,\sum\limits_{h,g}\frac{1}{2}\,\sum\limits_{a,b}{(-)^{a+b}~\frac{\theta[^a_b]^4\,\theta[^{a+h}_{b+g}]^8}{\eta^{12}}
}\cdot\frac{1}{2^2}\sum\limits_{\bar{a},\bar{b}}{\frac{\bar\theta[^{\bar{a}}_{\bar{b}}]^{16}\,\bar\theta[^{\bar{a}+h}_{\bar{b}+g}]^8}{\bar\eta^{24}}
}~.
\ee
The contributions to the partition function are organized in this case as:
$$
    \frac{1}{2}\sum\limits_{g}{Z_{(+,+)}[^0_g]\bar{Z}_{(+,+)}[^0_g]}=\left(V_{8}O_{16}-S_{8}C_{16}\right)\times\overline{\left(O_{32}O_{16}+C_{32}C_{16}\right)}=8\times\overline{\left(O_{32}O_{16}+C_{32}C_{16}\right)}
$$
$$
    \frac{1}{2}\sum\limits_{g}{Z_{(+,-)}[^0_g]\bar{Z}_{(+,-)}[^0_g]}=\left(O_{8}V_{16}-C_{8}S_{16}\right)\times\overline{\left(V_{32}V_{16}+S_{32}S_{16}\right)}=16\times\overline{\left(V_{32}V_{16}+S_{32}S_{16}\right)}
$$
$$
    \frac{1}{2}\sum\limits_{g}{Z_{(+,+)}[^1_g]\bar{Z}_{(+,+)}[^1_g]}=\left(V_{8}C_{16}-S_{8}O_{16}\right)\times\overline{\left(O_{32}C_{16}+C_{32}O_{16}\right)}=-8\times\overline{\left(O_{32}C_{16}+C_{32}O_{16}\right)}
$$
\be
    \frac{1}{2}\sum\limits_{g}{Z_{(+,-)}[^1_g]\bar{Z}_{(+,-)}[^1_g]}=\left(O_{8}S_{16}-C_{8}V_{16}\right)\times\overline{\left(\,V_{32}S_{16}+S_{32}V_{16}\right)\,}=0\times\overline{\left(\,V_{32}S_{16}+S_{32}V_{16}\right)\,}
\ee
The massless states of the shifted model are the following:
$$
    V_{8}O_{16}\overline{O_{32}O_{16}}~:8\times\overline{616}~~\textrm{states}~\chi_{-\frac{1}{2}}^{1\ldots
8}\otimes\{\bar\phi_{-\frac{1}{2}}^a\bar\phi_{-\frac{1}{2}}^b\oplus\bar\phi_{-\frac{1}{2}}^I\bar\phi_{-\frac{1}{2}}^J\}
$$
$$
    O_{8}V_{16}\overline{V_{32}V_{16}}~:~16\times\overline{512}~~\textrm{states}~\{y_{-\frac{1}{2}}^{1\ldots
8}\oplus w_{-\frac{1}{2}}^{1\ldots
8}\}\otimes\bar\phi_{-\frac{1}{2}}^a\bar\phi_{-\frac{1}{2}}^I
$$
$$
    -S_{8}O_{16}\overline{O_{32}C_{16}}~:~(-8)\times\overline{128}~~\textrm{states}~Sp\{\chi^{1\ldots
8}\}_{-}\otimes Sp\{\bar\phi^I\}_{+}
$$
We notice the appearance of $8$ massless fermions from
$-S_{8}O_{16}\overline{O_{32}C_{16}}$. Adding together the massless contributions we
find $n=n(b)-n(f)=12096,~n(f)=1024$, while the tachyonic
states in $V_{8}O_{16}\overline{O_{32}O_{16}}$ give the pole coefficient $m=8$. Therefore, the partition function of the $\Z_2$-shifted model is found to be:
\be
    Z_{het}^{B_s}=12096+8\,\left[\,\bar{j}(\bar\tau)-744\,\right]~.
\ee

\subsubsection{Heterotic $\Z_2$-shifted $MSDS$ with $SO(16)\times SO(16)\times E_8$}
A variation of the previous model can be obtained by further breaking $SO(32)$ to
$SO(16)\times SO(16)$. The breaking sets in this case are
$G=\{\bar\chi^{1\ldots 32}\}$ and $B_s=\{y^{1\ldots 8}w^{1\ldots 8}|\bar\chi^{33\ldots
48}\}$ and the corresponding partition function is:
\be
    Z_{het}^{G,B_s} =
\frac{1}{2}\,\sum\limits_{h,g}\frac{1}{2}\,\sum\limits_{a,b}{(-)^{a+b}~\frac{\theta[^a_b]^4\,\theta[^{a+h}_{b+g}]^8}{\eta^{12}}
}\cdot\frac{1}{2^2}\sum\limits_{\bar{a},\bar{b}\gamma,\delta}{\frac{\bar\theta[^{\bar{a}}_{\bar{b}}]^{8}\,\bar\theta[^{\bar{a}+h}_{\bar{b}+g}]^8\,\bar\theta[^\gamma_\delta]^8}{\bar\eta^{24}}
}~.
\ee
A similar analysis shows the presence of massless bosons and fermions with
$n=n(b)-n(f)=5952$,
$n(f)=2048$ and a pole coefficient $m=8$. The partition function is then:
\be
    Z_{heterotic}=5952+8\,\left[\,\bar{j}(\bar\tau)-744\,\right]~.
\ee

\subsection{Heterotic $MSDS$ $\Z_2\times\Z_2$-orbifolds }
There is a plethora of reduced $MSDS$ orbifold vacua in the heterotic framework. The
classification rules will be given in the next section where the (left-moving)
holomorphic $MSDS$ constraints will be derived.
Here we present a typical example that will be used as a representative paradigm in
Section 6 concerning
our discussion about the geometrical interpretation of the marginally deformed
$MSDS$-vacua. In fact, in the limit of large marginal deformations an effective four-dimensional
space-time will emerge. Aspiring to the construction of semi-realistic
four-dimensional heterotic chiral models, with $SO(10)\times U(1)^3\times SO(16)$
as right-moving gauge group,
we choose the basis set of the representative $MSDS$-vacuum to be:
$$
H_L=\{\chi^{1\ldots8},y^{1\ldots 8},w^{1\ldots 8}\},~~
H_R=\{\bar{y}^{1\ldots 8},\bar{w}^{1\ldots
8},\bar\eta^1,\bar\eta^2,\bar\eta^3,\bar\psi^{1\ldots 5},\bar{\phi}^{1\ldots 8}\},
$$
$$
G=\{\bar{y}^{1\ldots 8},\bar{w}^{1\ldots 8}\},~~
z=\{\bar\phi^{1\ldots 8}\}
$$
\be
b_1=\{\chi^{3,4,5,6},y^{3,4,5,6}|\bar{y}^{3,4,5,6},\bar\eta^1,\bar\psi^{1\ldots
5}\},~~~
b_2=\{\chi^{1,2,5,6},y^{1,2,5,6}|\bar{y}^{1,2,5,6},\bar\eta^2,\bar\psi^{1\ldots 5}\}.
\ee
In the above $\chi^I,y^I,w^I,\bar{y}^I,\bar{w}^I$ are considered to be real fermions
while $\bar{\eta}^{1,2,3},\bar\psi^{1\ldots 5}$ and $\bar\phi^{1\ldots 8}$ are
complex.
The holomorphic part of the partition function is:
\be
    Z[^{h_1~h_2}_{g_1~g_2}]=\frac{1}{2}\sum\limits_{a,b}{(-)^{a+b}\,\frac{\theta[^a_b]^6\,\theta[^{a+h_1}_{b+g_1}]^2\,\theta[^{a+h_2}_{b+g_2}]^2\,
    \theta[^{a-h_1-h_2}_{b-g_1-g_2}]^2}{\eta^{12}}}~,
\ee
whereas the anti-holomorphic part is:
\be
    \bar{Z}[^{h_1~h_2}_{g_1~g_2}]=\frac{1}{2^3\,\bar\eta^{24}}\sum\limits_{\gamma,\delta}
    {\bar\theta[^\gamma_\delta]^5\,\bar\theta[^{\gamma+h_1}_{\delta+g_1}]\,
    \bar\theta[^{\gamma+h_2}_{\delta+g_2}]\,
    \bar\theta[^{\gamma-h_1-h_2}_{\delta-g_1-g_2}]}\,\sum\limits_{\epsilon,\zeta}
    {\bar\theta[^\epsilon_\zeta]^5\,\bar\theta[^{\epsilon+h_1}_{\zeta+g_1}]\,
    \bar\theta[^{\epsilon+h_2}_{\zeta+g_2}]}
\,\bar\theta[^{\epsilon-h_1-h_2}_{\zeta-g_1-g_2}]\,\sum\limits_{\bar a,\bar
b}{\bar\theta[^{\bar a}_{\bar b}]^8}~.
\ee
The full partition function can be written in a conventional {\it shifted and
twisted} ``$\Gamma_{8,8}$-lattice form" \cite{ShifftedLat,R-Shifted} as follows :
$$
 Z_{\Z_2\times\Z_2}=\frac{1}{2^6\,\eta^{12}\bar\eta^{24}}\sum\limits_{a,b,\gamma,\delta,h_i,g_i}{(-)^{a+b}\,\theta[^a_b]\,\theta[^{a+h_1}_{b+g_1}]\,\theta[^{a+h_2}_{b+g_2}]\,
\theta[^{a-h_1-h_2}_{b-g_1-g_2}]}~\times
$$
\be
\label{G88form}
\times
\,\Gamma_{8,8}\left[^{a~,~\gamma}_{b~,~\delta}
\right|\left.^{h_i}_{g_i}\right]\,\sum\limits_{\epsilon,\zeta}{\bar\theta[^\epsilon_\zeta]^5\,\bar\theta[^{\epsilon+h_1}_{\zeta+g_1}]\,\bar\theta[^{\epsilon+h_2}_{\zeta+g_2}]\, \bar\theta[^{\epsilon-h_1-h_2}_{\zeta-g_1-g_2}]}\,\sum\limits_{\bar{a},\bar{b}}{\bar\theta[^{\bar{a}}_{\bar{b}}]^8}\,,
\ee
where $\Gamma_{8,8} \left[^{a~,~\gamma}_{b~,~\delta}\right|\left.^{h_i}_{g_i}\right]$
 indicates the contribution  of the eight  fermionized coordinates
$\{y^I,\omega^I\,|\,\bar y^I,\bar\omega^I\}$.
The $MSDS$-structure of the holomorphic side has already been studied in Section 3,
where
$\Z_2\times\Z_2$ Type-II models were considered. Thus, the heterotic model under
consideration also possesses $MSDS$-structure. Specifically, the partition function
in this
representative example is found to be:
\be
Z=12\bar{j}(\bar\tau)=12\,\times744+12\,\left[\,\bar{j}(\bar\tau)-744\,\right]~.
\ee
Inserting in the above representative model all possible discrete torsion
coefficients permitted by
the fermionic construction, a plethora of $MSDS$ Heterotic models can be obtained.
The resulting models will in general exhibit different
bosonic and fermionic massless spectra in different representations of the chiral
(right-moving)
gauge group  $SO(10)\times U(1)^3\times SO(16)$, similarly to the four-dimensional
supersymmetric chiral models of ref.\,\cite{FKR}. The generic property of all
those models is that their partition function will be always of the form of Eq.
(\ref{genericHet}).


\section{Classification of fermionic $MSDS$ Vacua}

In the previous section we presented various examples of $\Z_2$-orbifolds of the
original Type II and Heterotic $MSDS$-vacua and showed that the $MSDS$-structure
is inherited by
the ``daughter" models we presented. There it was mentioned that the $MSDS$-structure
of the spectrum results from a consistent truncation of the original chiral
superconformal algebra. In this section we elaborate on the necessary conditions for
such a truncation to be consistent and obtain a simple set of rules that permit the
construction of all fermionic $MSDS$-vacua.

$~$\\
As $MSDS$-structure has been seen to be a chiral property, it is sufficient for now
to concentrate on the holomorphic side. Once $MSDS$-symmetry in the sense of
spectral-flow has been secured on the holomorphic side, one may couple it to an
arbitrary anti-holomorphic part, respecting only constraints of modular invariance.
The resulting model will necessarily have $MSDS$-structure, since the partition
function in that case can easily be shown to equal at most a constant plus a
possible term proportional to the Klein $j$-invariant. After projecting out the
unphysical states, one remains with the surviving constant contribution of the
massless states only.

$~$\\
To motivate the consistency conditions we are about to derive, we consider the simple
$\Z_2$-orbifold already examined in the previous sections. The holomorphic part of
the partition function is:
\be
    (-)^{a+b+hg}~\frac{\theta[^a_b]^8\,\theta[^{a+h}_{b+g}]^4}{\eta^{12}}~,
\ee
where we explicitly added the phase $(-)^{hg}$ to make the holomorphic part modular
invariant by itself. The contributions to the untwisted sector are
$O_{16}V_8-C_{16}S_8$ and $V_{16}O_8-S_{16}C_8$. We first note the existence of a
$\Z_2$-invariant spectral-flow operator:
\be
    \textbf{C}=e^{\frac{1}{2}(\Phi-iH_0)}\, {C}_{16}{C}_8~.
\ee
Such an operator must map massive boson states to massive fermion states in the
untwisted sector because the latter is simply a projection of the initial
$MSDS$-vacuum. We notice that the dressed spectral-flow operator $\textbf{C}(z)$ has
conformal
weight $\Delta_{\bf C}=1$ and, thus, it effectively acts as a current:
\be
    j_{\rm{MSDS}}(z)~\equiv ~\textbf{C}(z)~.
\ee
This spectral-flow is responsible for the isomorphism that maps the massive tower of
states $O_{16}V_8$ into $C_{16}S_8$, leaving only the massless states invariant. For
this mapping to exist, as in the case of ordinary supersymmetry, we must be able to
define a BRST-invariant charge $Q_{\rm{MSDS}}$ :
\be
    Q_{\rm{MSDS}}=\oint{\frac{dz}{2\pi i}~ j_{\rm{MSDS}}(z)}=\oint{\frac{dz}{2\pi
i}~ e^{\frac{1}{2}(\Phi-iH_0)}\,C_{16}C_{8}(z)}~,
\ee
with a well-defined action on the states of the spectrum. We will refer to this
operator as the \emph{MSDS charge}. The problem of
classifying \emph{MSDS}-vacua is, therefore, twofold. One must first examine under
what conditions the action of the \emph{MSDS}-charge is well-defined on the spectrum
of states and, secondly, to ensure that massless states are annihilated by its action.

$~$\\
 The states of the vacuum representation contributing to the $O_{16}$-character can
be split into those generated by the identity operator $\textbf{1}_{16}$ (with
conformal weight $\Delta_{\textbf{1}}=0$), as well as by the those in the adjoint
$\textbf{Adj}_{16}(z)=(\hat\chi\hat\chi)_{16}$ (with weight
$\Delta_{\textbf{Adj}}=1$). In what follows, we consider the adjoint as part
of the affine descendants of the identity operator. The fusion relation shows that massless states
do not transform, since in this case the \emph{MSDS}-charge vanishes:
\be\label{fusion1}
    j_{\rm{MSDS}}(z)\cdot\left(\textbf{1}_{16}\textbf{V}_8\right)(w) \sim
\textbf{C}_{16}\textbf{S}_8(w)~.
\ee
On the other hand, the massive states do transform:
\be\label{sf1}
    j_{\rm{MSDS}}(z)\cdot\left(\textbf{Adj}_{16}\textbf{V}_8\right)(w) \sim
\frac{\textbf{C}_{16}\textbf{S}_8(w)}{z-w}~.
\ee
The same \emph{MSDS}-mapping would, of course, be true for the descendant states
generated by $(\textbf{1}_{16}\textbf{V}_{(\textbf{1}),\,8})$:
\be\label{sf2}
 j_{\rm{MSDS}}(z)\cdot\left(\textbf{1}_{16}\textbf{V}_{(\textbf{1}),\,8}\right)(w) \sim
\frac{\textbf{C}_{16}\textbf{S}_8(w)}{z-w}~,
\ee
 where
\be
\textbf{V}_{(\textbf{1})}(z)~\equiv~ e^{-\Phi}~\left(~\partial\chi~+~\hat\chi\hat\chi\hat\chi~\right)
\ee
is the first descendant operator of $\textbf{V}_{(\textbf{0})}(z)$.

$~$\\
The spectral-flow relations (\ref{fusion1}), (\ref{sf1}) and (\ref{sf2}) are responsible for the
isomorphism that maps the massive tower of
states $O_{16}V_8$ into $C_{16}S_8$, leaving only the massless states invariant.
This can be seen explicitly by considering the action of the \emph{MSDS}-charge on
the massless and massive states mentioned above.
The difference of characters is, therefore, completely determined by the excess in
massless states, giving:
\be
O_{16}V_8 - C_{16}S_8 = 8-0 =8~,
\ee
since only $O_{16}V_8$ contributes to the massless spectrum. Now consider the action
of the orbifold. Under the $\Z_2$-twist the vertex operators charged under $SO(8)$
transform as $O\leftrightarrow C$ and $V\leftrightarrow S$, respecting parity.
In order to preserve $MSDS$ symmetry in the twisted sector we need to ensure that
the twisted states have similar fusion rules:
\be
    j_{\rm{MSDS}}(z)\cdot\left(\textbf{1}_{16}\textbf{S}_8\right)(w) \sim
\textbf{C}_{16}\textbf{V}_8(w)
\ee
\be \label{sf3}
    j_{\rm{MSDS}}(z)\cdot\left(\textbf{Adj}_{16}\textbf{S}_8\right)(w) \sim
\frac{\textbf{C}_{16}\textbf{V}_8(w)}{z-w}~,
\ee
so that, again, the massless states remain invariant, whereas the massive ones
transform. A similar fusion rule is responsible for the mapping of the remaining
massive descendants of $\textbf{1}_{16}\textbf{S}_8$ into
$\textbf{C}_{16}\textbf{V}_8$.
 The spectral-flow is inherited to the twisted sector and one obtains the
twisted character formula:
\be
    -O_{16}S_8 + C_{16}V_8 = -8~.
\ee
On the other hand, one could also perform the inverse mapping:
\be
    j_{\rm{MSDS}}(z)\cdot\left(\textbf{C}_{16}\textbf{S}_8\right)(w) \sim
\frac{\textbf{1}_{16}\textbf{V}_8(w)}{(z-w)^2}+\frac{\textbf{Adj}_{16}\textbf{V}_8(w)+\textbf{1}_{16}\textbf{V}_{\textbf{(1)},8}(w)}{z-w}~.
\ee
This is in exact agreement with our previous results, showing the mapping of the
spinorial primary states generated by $\textbf{S}(z)$ into the descendant bosonic
states in
$\textbf{Adj}_{16}\textbf{V}_8(z)$ and
$\textbf{1}_{16}\textbf{V}_{\textbf{(1)},8}(z)$ that lie at the same mass level.
This explicitly demonstrates that the
\emph{MSDS}-mapping:
\be
    \{~~\textbf{S}(0)|0\rangle~~\} \leftrightarrow \{~~
\textbf{Des}_{\textbf{(1)}}\textbf{[} \textbf{1}_{16}\textbf{V}_{8} \textbf{]}~|0\rangle
~~\}
\ee
is indeed bijective, as it should. For convenience, we denote as:
\be
    \{~~\textbf{Des}_{\textbf{(1)}}\textbf{[} \textbf{1}_{16}\textbf{V}_{8}
\textbf{]}~|0\rangle ~~\} \equiv
\{~~\textbf{Adj}_{16}\textbf{V}_8(0)|0\rangle~~\oplus~~
\textbf{1}_{16}\textbf{V}_{\textbf{(1)},8}(0)|0\rangle~~\}
\ee
the set of states generated by the descendant operators of
$\textbf{1}_{16}\textbf{V}_8$ at the next mass level. These descendant operators, in
turn, generate the remaining
tower of massive states of the vectorial representation by the repeated action of
the `covariant-like derivative' $\mathcal{D}\equiv \partial+\hat\chi\hat\chi$.

$~$\\
It is now important to notice that the spectral-flow is valid in the twisted sector
only for {\it specific
choices of the breaking sets}. Consider introducing a breaking set $b$ twisting
$n_L(b)$ left-moving fermions and an appropriate number of right-moving fermions in
such a way that modular invariance constraints are satisfied. The relevant
characters will now transform under $SO(24-n_L(b))\times SO(n_L(b))$. The spectral
flow is always valid in the untwisted sector:
\be
j_{\rm{MSDS}}(z)\cdot\left(\textbf{1}_{24-n_L}\textbf{V}_{n_L}\right)(w)
\sim \textbf{C}_{24-n_L}\textbf{S}_{n_L}(w)
\ee
\be
j_{\rm{MSDS}}(z)\cdot\textbf{Des}_{\textbf{(1)}}\textbf{[}\textbf{1}_{24-n_L}\textbf{V}_{n_L}\textbf{]}(w)
\sim \frac{\textbf{C}_{24-n_L}\textbf{S}_{n_L}(w)}{z-w}~,
\ee
where the symbol $\textbf{Des}_{\textbf{(q)}}\textbf{[}\textbf{A}\textbf{]}$ stands
for the descendants of $\textbf{A}$ starting from the $q$-th next mass level after
that
of $\textbf{A}$.

$~$\\
We now see that this fusion rule {\it is not preserved by the twist for generic
$n_L(b)$}.
The corresponding fusion rule in the twisted sector is:
\be
j_{\rm{MSDS}}(z)\cdot\textbf{Des}_{\textbf{(q)}}
\textbf{[}\textbf{1}_{24-n_L}\textbf{S}_{n_L}\textbf{]}(w)
\sim
\sum\limits_{q'=0}{\frac{\textbf{Des}_{\textbf{(q}'\textbf{)}}
\textbf{[}\textbf{C}_{24-n_L}\textbf{V}_{n_L}(w)\textbf{]}}{\left(z-w\right)^{n_L/8-1+q-q'}}}~,
\ee
where $q,q'\in\{0,1,2,3,\ldots\}$. The cases $q=0$, $q'=0$ correspond to the original
(primary) operators $\textbf{1}_{24-n_L}\textbf{S}_{n_L}$ and
$\textbf{C}_{24-n_L}\textbf{V}_{n_L}$,
respectively. Clearly the spectral-flow algebra is preserved under the twist only
for choices $b$ of breaking sets satisfying $n_L(b)=0\,(\textrm{mod}\,8)$, otherwise
the fusion OPE contains cuts and the action of the current on the vertex
operators is non-local. In the latter case, the \emph{MSDS}-charge $Q_{\rm{MSDS}}$ cannot
be defined, implying the absence of mapping from massive bosonic to fermionic
states.
Assuming now that the \emph{MSDS}-charge is well-defined, the mapping can only occur if
there appears a simple pole in the OPE, implying:
\be
  \frac{1}{8}\,n_L+q-q'=2
\ee
for some $q'\geq 0$.
It is easy to see that this condition is only violated by the massless states, which
correspond precisely to $n_L=8$ and $q=0$ and, therefore, they do not transform.

$~$\\
We are now ready to formulate the analogous argument for a generic $(\Z_2)^N$-twist.
Generically, acting with the \emph{MSDS}-charge on a state
$|\mathcal{A}\rangle$, created by the local operator $\mathcal{A}(w)$ with weight
$\Delta_{\mathcal{A}}$, one produces a state $|\mathcal{B}\rangle$ that is created
by an operator $\mathcal{B}(w)$ with weight $\Delta_{\mathcal{B}}$. Taking into
account the possible descendants that can appear, the fusion relation is:
\be
    j_{\textrm{MSDS}}(z)\cdot\textbf{Des}_{(q)}\textbf{[}\mathcal{A}\textbf{]}(w)\sim
\sum\limits_{q'=0}{\frac{
\textbf{Des}_{(q')}\textbf{[}\mathcal{B}\textbf{]}(w)
}{(z-w)^{\Delta_{\mathcal{A}}-\Delta_{\mathcal{B}}+q-q'+1}}   }~.
\ee
Since only simple poles give a spectral-flow mapping\footnote{This means that only
states of the same mass level can map into each other, as should be expected.}, the
condition for the existence of the \emph{MSDS}-mapping between the two states is:
\be\label{condition}
    \Delta_{\mathcal{A}}-\Delta_{\mathcal{B}} = q'-q
~~,~~\textrm{where}~q,q'\in\{0,1,2,\ldots\}~.
\ee
The case $\Delta_{\mathcal{A}}=\Delta_{\mathcal{B}}$ corresponds to a `primary'
$(q=0)$ operator $\mathcal{A}$ being mapped into another `primary' $(q'=0)$ operator
$\mathcal{B}$.
The case $\Delta_{\mathcal{A}}\neq\Delta_{\mathcal{B}}$ implies that the dominant
contributions to the OPE come with higher-order poles and, therefore, only their
descendants
appearing with simple poles eventually contribute to the mapping. Supposing that the
\emph{MSDS}-charge is well-defined for a particular model, the massless states are
the
only ones that always violate (\ref{condition}) for any $q'\geq 0$, as they correspond to
$\Delta_{\mathcal{A}}=1/2$ and $q=0$.

The \emph{MSDS}-charge is, therefore, well-defined on all states provided that
$\Delta_{\mathcal{A}}-\Delta_{\mathcal{B}}\in\Z$ for any primary states
$\mathcal{A},\mathcal{B}$. While this is trivially satisfied in the untwisted
sector, in the twisted sector it imposes powerful constraints on the form
of permitted breaking sets. To show this, consider the most general contribution
to the untwisted sector:
\be
\mathcal{A}(w)=\prod\limits_{n_i}^{n}{\textbf{V}_{n_i}}\prod\limits_{m_j}^{m}{\textbf{O}_{m_j}}(w)~,
\ee
which has conformal weight $\Delta_{\mathcal{A}}=n/2$. Of course, consistency
requires that:
\be
\sum\limits_{n_i}^n{n_i}+\sum\limits_{m_j}^m{m_j}= 24~.
\ee
Acting with the spectral flow operator $j_{\rm{MSDS}}(z)$  we obtain another
operator:
\be
\mathcal{B}(w)=\prod\limits_{n_i}^{n}{\textbf{S}_{n_i}}\prod\limits_{m_j}^{m}{\textbf{C}_{m_j}}(w)~,
\ee
with conformal weight $\Delta_{\mathcal{B}}=3/2$.
The condition (\ref{condition}) for the existence of a mapping between the two
states is :
\be
\label{trivialcondition}
\frac{n-3}{2} = q'-q~.
\ee
Since the $GGSO$-projection of the models under consideration forces the overall
parity to be negative, $n$ will always be odd so that
$\Delta_{\mathcal{A}}-\Delta_{\mathcal{B}}$ is always an integer. It
is important to note that only the massless states violate this condition for all
$q'\geq 0$, since they
correspond to $q=0$ and $n=1$. All other cases correspond to massive states and transform, in agreement to our previous considerations.

$~$\\
 We now proceed to impose condition (\ref{condition}) on the twisted sector. The
original operator becomes, under a generic $\Z_2$-like twist:
\be
\mathcal{A}'(z)=\prod\limits_{n_i}^{n-a}{\textbf{V}_{n_i}}\prod\limits_{n_i}^{a}{\textbf{S}_{n_i}}\prod\limits_{m_j}^{m-b}{\textbf{O}_{m_j}}\prod\limits_{m_j}^{b}{\textbf{C}_{m_j}}(w)~,
\ee
with conformal weight:
\be
\Delta_{\mathcal{A}'}=
\frac{n-a}{2}+\sum\limits_{n_i}^{a}{\frac{n_i}{16}}+\sum\limits_{m_j}^{b}{\frac{m_j}{16}}~.
\ee
The action of $j_{\rm{MSDS}}(z)$ on the `twisted' operator produces:
\be
\mathcal{B}'(z)=\prod\limits_{n_i}^{n-a}{\textbf{S}_{n_i}}\prod\limits_{n_i}^{a}{\textbf{V}_{n_i}}\prod\limits_{m_j}^{m-b}{\textbf{C}_{m_j}}\prod\limits_{m_j}^{b}{\textbf{O}_{m_j}}(w)~,
\ee
with conformal weight:
\be
\Delta_{\mathcal{B}'}=
\frac{a}{2}+\sum\limits_{n_i}^{n-a}{\frac{n_i}{16}}+\sum\limits_{m_j}^{m-b}{\frac{m_j}{16}}~.
\ee
In this case, condition (\ref{condition}) becomes simply:
\be
\label{condition2}
    \frac{1}{8}\left(\sum\limits_{n_i}^a{n_i}+\sum\limits_{m_j}^b{m_j}\right)+\frac{n-3}{2}-a
= q'-q~.
\ee
To ensure that the above is indeed an integer we require:
\be
    \sum\limits_{n_i}^a{n_i}+\sum\limits_{m_j}^{b}{m_j}=0\,(\textrm{mod}\,8)~,
\ee
which forces the left-moving twisted fermions to appear in multiples of eight:
\be\label{mod8rule}
    n_L(b_i)=0\,(\textrm{mod}\,8)
\ee
for any element $b_i$ in the basis of the parity group $\Xi$. Note that the
mapping condition (\ref{condition2}) is violated at precisely the massless cases, as would be expected.
Indeed, massless states in the twisted sector correspond to:
\be
   a=n~~,~~q=0
\ee
and
\be
    \frac{1}{16}\left(\sum\limits_{n_i}^a{n_i}+\sum\limits_{m_j}^{b}{m_j}\right)=\frac{1}{2}~,
\ee
for which  (\ref{condition2}) is violated for any $q'\geq 0$, since $n$ is
odd by the \emph{GGSO}-projections.
In fact, (\ref{mod8rule}) together with overall modular invariance of the full
partition function automatically imply the following two conditions on
the left-moving degrees of freedom of basis elements:
\be
    n_L(b_i\cap b_j)=0\,(\textrm{mod}\,4)~~,~~n_L(b_i\cap b_j\cap
b_k)=0\,(\textrm{mod}\,2)~.
\ee
The above constraints imply that the basis elements in $\Xi$, if effectively
truncated to their left-moving parts only, would still generate a holomorphic
modular invariant partition function. It is clear that this holomorphic partition
function can only equal a constant integer, since no term linear in the Klein
$j$-invariant can be constructed by quantizing only $24$ real
fermions\footnote{Indeed, to create a Klein $j(\tau)$-function one would need $48$
real fermions in $2$ space-time dimensions. This can only happen in the absence of
worldsheet supersymmetry, as in the heterotic case.}.

$~$\\
We finally summarize the above results into a classification theorem that permits the
$\Z_2^N$-orbifold construction of all fermionic $MSDS$-vacua in 2 flat spacetime dimensions:

\paragraph{Theorem:} For any choice of parity group $\Xi$ containing $n_L(F)=24$
free left-moving fermions whose basis elements satisfy, in addition to the usual
modular invariance constraints, the \emph{holomorphic} constraint:
\be
\nonumber\label{holConstr}
n_L(b_i)=0\,(\textrm{mod}\,8)~,
\ee
and whose basis elements $b_i\in\Xi$ preserve the global definition\footnote{This
requirement can be easily imposed by requiring the sum of all twist/shift indices to
vanish $\sum_i{h_i}=0$, similarly to the case of ordinary supersymmetry.} of the
spectral-flow operator $j_{\rm{MSDS}}(z)$, the resulting fermionic model has
$MSDS$-structure and its partition function, depending on the anti-holomorphic
structure, will at most equal a constant, plus an additional linear antiholomorphic $j$-invariant
term in the Heterotic case.

$~$\\
 In Type-II and Heterotic theories the anti-holomorphic
constraint $n_R(b_i)=0\,(\textrm{mod}\,8)$ for the right-movers is automatically
satisfied once (\ref{holConstr}) is imposed on the left-movers, because of modular
invariance. The $MSDS$-structure, thus, automatically appears in the anti-holomorphic side of
Type II theories as well and the partition function is simply equal to a constant.

$~$\\
On the other hand, in Heterotic models where there is no anti-holomorphic
spectral-flow operator to guarantee $MSDS$-structure in the right-moving side, an
additional Klein $j$-function is generated. The latter, in turn, participates in the massless
contribution while its ``spurious" massive terms will eventually give zero
contribution to the integrated partition function. These remarks permit the
construction of all real fermionic Type II and Heterotic $MSDS$-Vacua.

\section{Marginal deformations of  $RMSDS$ Orbifold Vacua }
 The   initial  $2d$ $MSDS$ string vacua proposed in ref.~\cite{MSDS} are non-geometrical in terms of  the internal compactifed
 space but are rather characterized by the non-abelian  gauge group $H_L\times H_R$.
In the massless spectrum there are scalar
 bosons  $M_{I_L,J_R},~I_L=1,2,...,d_L,~J_R=1,2,...,d_R~$, parametrizing  the manifold given in
 Eq.~(\ref{manifK}).
As was already mentioned in the introduction,  because of the non-abelian structure of $H_L\times H_R$, the $MSDS$ vacua admit marginal deformations (flat directions) associated with the Cartan
sub-algebra $U(1)^{r_L}\times U(1)^{r_R}$, with $r_L$ and $r_R$ being the ranks of  $H_L$ and  $H_R$ respectively as in Eq.~(\ref{manifM}).

$~$\\
Ultimately, the $M_{IJ}$-deformation
parameters are connected with an  {\it ``induced effective higher dimensional space
geometry" } in  the {\it large $M_{IJ}$-deformation limit} (e.g. when the
$MSDS$-vacua are strongly-deformed). In this limit one recovers the geometrical
field theory description in terms of  an effective  ``higher dimensional"
conventional  superstring theory in which space-time supersymmetry is
{\it spontaneously broken}  by ``geometrical" and ``thermal" fluxes. This
fundamental generic  property of the  deformed $MSDS$-vacua  suggests the following
  {\it Cosmological Conjecture} formulated in ref.~\cite{MSDS}:
\begin{itemize}
\item
  The $MSDS$-vacua, (or most likely their less symmetric orbifold reductions, such as those considered here) are potential candidates to describe {\it the early non-singular phase of a stringy cosmological universe}.
  \item
 The deformation moduli $M_{IJ}\rightarrow M_{IJ}(t)$ are subject to cosmological evolution and as such, they eventually acquire time dependence. Once  the $M_{IJ}(t)$ become sufficiently large (modulo $S,T,U$-dualities), an effective field theory description  emerges along with an induced ``space-time geometry" of an {\it effective higher dimensional  space-time}. The relevant degrees of freedom and interactions will be then described by the effective ``no-scale" supergravity  theories \cite{Noscale} of conventional superstrings \cite{StringyNoscale}.
  \item
The $MSDS$-structure at the early cosmological times induces, in the large moduli limit, non-trivial ``geometrical" fluxes \cite{GeoFluxes,GravFluxes, Cosmo-RT-Shifted} which,  in the language of the effective supergravity, give rise to a spontaneous breaking of supersymmetry \cite{ShifftedLat,R-Shifted} and to a finite temperature description of the effective theory \cite{AKADK,GravFluxes, Cosmo-RT-Shifted}.
 \end{itemize}
In this respect one may consider  $RMSDS$-models as the most (semi-) realistic candidate vacua able to describe the  ``early non-singular phase of our Universe", being free of any
initial ``general relativity-like'' or ``Hagedorn-like'' stringy singularities.

 $~$\\
The moduli space of the $RMSDS$ orbifolds obviously containts a subspace of would-be {\it geometrical $M_{IJ}$-deformations}, associated with the conventional supersymmetric (freely acting)
orbifolds. For instance, in the representative $\Z_2\times \Z_2$
Heterotic example defined in Eq.~(\ref{G88form}), the
shifted lattice  $\Gamma_{8,8}
\left[^{a~,~\gamma}_{b~,~\delta}\right|\left.^{h_i}_{g_i}\right]$ is
also twisted by $\Z_2\times \Z_2$ and so, the initial deformation
space is reduced to:
\be
\Z_2\times \Z_2 :~  {SO(8,8)\over
SO(8)\times SO(8)}~\longrightarrow~ {SO(4,4)\over SO(4)\times
SO(4)} \times{SO(2,2)\over SO(2)\times SO(2)}\times {SO(2,2)\over
SO(2)\times SO(2)}~.
\ee
Assuming very large deformations in
the $(2,2)$ sub-space of $SO(4,4)$, a four-dimensional flat space-time is
generated, together with a six-dimensional compact space described
by $T^6\over \Z_2\times \Z_2$. This class of models is then connected with the semi-realistic $N=1$ chiral supersymmetric
models based on the $SO(10)$ unified gauge group which were
classified in ref.~\cite{FKR}. Furthermore, the $RMSDS$-deformed
models in this class provide vacua with $N=1$ supersymmetry, spontaneously broken by very specific geometrical fluxes! This remarkable
property follows from the fact that the initial $\Gamma_{8,8} \left[^{a~,~\gamma}_{b~,~\delta}\right|\left.^{h_i}_{g_i}\right]$ lattice is shifted by a set of well-defined
$R$-symmetry charges \cite{ShifftedLat,R-Shifted}, as dictated by the non-deformed $RMSDS$ vacua.
 Generically, for  large but not infinitely large deformations, the obtained vacua are those of  ``spontaneously broken supersymmetric vacua in the presence of geometrical fluxes'' \cite{ GeoFluxes}, studied in detail in refs \cite{ShifftedLat, R-Shifted}. Notice also that in the Euclidian version some of the models correspond to ``thermal stringy vacua" in the presence of non-trivial left-right asymmetric  ``gravito-magnetic fluxes" studied recently in refs \cite{GravFluxes, Cosmo-RT-Shifted}.
 The would-be ``initial" classical singularity of general relativity as well as the stringy Hagedorn-like singularities are both resolved by these fluxes !

\section{Conclusions}
The existence of a new massive boson-fermion degeneracy symmetry is shown by explicit orbifold constructions  in  Type II  and Heterotic string theories. In all constructions, the target space-time is 2-dimensional and the spectrum consists of massless bosonic degrees of freedom as well as of massless fermionic ones
with $(n(b)-n(f))\ne 0$.  All massive boson and fermion degrees of freedom exhibit  Massive Spectrum Degeneracy Symmetry ($MSDS$). This remarkable property follows from the
modular properties between the Vector ($V$), Spinor ($S$) and Anti-Spinor ($C$) characters of the affine $G\subset SO(24)$ algebra, twisted by the $\Z_2$-orbifolds that are formulated algebraically in terms of {\it twisted} $\theta^{12}$-identities.

$~$\\
A new chiral $\mathcal{N}=1$ superconformal algebra is proposed based on the usual $\mathcal{N}=1$ super-Virasoro operators  $T_B~(h_B=2)$ and  $T_F ~(h_F=3/2)$, together with
 $C ~(h_C=3/2)$ and $J^a ~(h_J=1)$, where $J^a$ are the currents of the semi-simple gauge group $H$ reduced by the orbifolds. The reduced massive boson-fermion
 degeneracy follows from a ``spectral flow" relation induced by the algebra $\{ T_B,~T_F, ~C, ~J^a \} $.
In this work we derived the necessary conditions leading to the classification {\it of all} fermionic $\Z_2^N$-orbifold constructions of vacua with $MSDS$-structure.
These classification rules are of main importance since the $RMSDS$-vacua are eventually related, via ``Cosmological Large Marginal Deformations", to some  effective
 ``four-dimensional" semi-realistic chiral  superstring vacua  with spacetime supersymmetry {\it spontaneously broken} by the $RMSDS$-induced ``geometric" and ``thermal" fluxes. The connection of $RMSDS$-vacua with ``gauged supergravity theories" is by now  transparent in the ``strongly deformed phase" via the induced geometrical fluxes of the effective higher-dimensional theories.
 It is, thus, strongly suggested that the  deformed  orbifold $RMSDS$-models be considered as the most (semi-)realistic candidate vacua able  to describe the  ``early non-singular phase of
 our Universe", free of initial ``general relativity-like" as well as of any ``Hagedorn-like" singularities.

 $~$\\
The observation that massless space-time fermions can appear in the twisted sectors of $RMSDS$-orbifold constructions hints at the possibility  of constructing field theories
 with unbroken $RMSDS$ and massless chiral fermions in higher than two dimensions, the case of $d=4$ dimensions being the most theoretically and phenomenologically appealing.
 Progress in that direction may produce interesting alternatives to the conventional supersymmetry approach, possibly even bypassing some of the well-known mathematical inconsistencies
 related to the hierarchy and to the cosmological constant problems.

$~$\\
Finally, a noticeable  property of 2-dimensional $RMSDS$-orbifold vacua is the {\it holomorphic  factorization} property of their partition function.  Although these theories have
non-trivial massive spectra, thanks to the $MSDS$ structure, all non-topological degrees of freedom are effectively washed out of the partition function! In this respect,
$RMSDS$-orbifold vacua realize  2d target-space conformal field theories with holomorphic factorization properties similar to those initially proposed by Witten
\cite{WittenHoloFactor} in connection with BTZ-black holes \cite{HennStrominger}. In this context, the 2d  $MSDS$-vacua (especially the Heterotic ones) are identified with
the boundary 2d conformal field theories of $AdS_3$ \cite{HennStrominger}. Following Witten's conjecture, the massive  bosonic spectrum is identified with the mass spectrum
of BTZ-black holes \cite{WittenHoloFactor}. The $MSDS$-theories, however, additionally suggest the existence of a fermionic ``massive supersymmetric" partner having the same
mass spectrum as the bosonic one !


\section*{Acknowledgements}

We are  grateful to C.~Bachas,  H.~Partouche, C.~Angelantonj and especially to N.~Toumbas  and J.~Troost for useful and
fruitful  discussions. I.F. would also like to thank F. Bourliot and C. Condeescu for several stimulating discussions.  This work  is partially supported by the EU contract
MRTN-CT-2004-005104
and the ANR (CNRS-USAR) contract  05-BLAN-0079-01.



\begin{thebibliography}{99}


\bibitem{GSW}
  M.~B.~Green, J.~H.~Schwarz and E.~Witten,
  ``Superstring Theory", Vol. 1 and Vol. 2:
{\it  Cambridge, UK: Univ. Pr.} ( 1987), Cambridge Monographs on Mathematical Physics.

  J.~Polchinski,
  ``String theory" Vol. 1 and Vol. 2: ``An introduction to the bosonic string,'' and
``Superstring theory and beyond,''
{\it  Cambridge, UK: Univ. Pr. (1998) 402 p} and {\it  Cambridge, UK: Univ. Pr.
(1998) 531 p}.


  E.~Kiritsis,
  ``Introduction to superstring theory,'' {\it Leuven Notes in Mathematical and
Theoretical Physics}, V.9: arXiv:hep-th/9709062,
  E.~Kiritsis,
  ``String theory in a nutshell,''
{\it  Princeton, USA: Univ. Pr. (2007) 588 p}.



\bibitem{CosmoTopologyChange}
  E.~Kiritsis and C.~Kounnas,
  ``Dynamical topology change, compactification and waves in a stringy early
  universe,''
  arXiv:hep-th/9407005.

  E.~Kiritsis and C.~Kounnas,
  ``Dynamical topology change in string theory,''
  Phys.\ Lett.\  B {\bf 331} (1994) 51
  [arXiv:hep-th/9404092].


  E.~Kiritsis and C.~Kounnas,
  ``Dynamical topology change, compactification and waves in string
  cosmology,''
  Nucl.\ Phys.\ Proc.\ Suppl.\  {\bf 41} (1995) 311
  [arXiv:gr-qc/9701005].

  E.~Kiritsis and C.~Kounnas,
  ``String gravity and cosmology: Some new ideas,''
  arXiv:gr-qc/9509017.



\bibitem{GV}
  M.~Gasperini and G.~Veneziano,
  ``Pre-big bang in string cosmology,''
  Astropart.\ Phys.\  {\bf 1} (1993) 317
  [arXiv:hep-th/9211021].

  M.~Gasperini, M.~Maggiore and G.~Veneziano,
  ``Towards a non-singular pre-big bang cosmology,''
  Nucl.\ Phys.\  B {\bf 494}, 315 (1997)
  [arXiv:hep-th/9611039].


\bibitem{BV}
  R.~H.~Brandenberger and C.~Vafa,
  ``Superstrings in the Early Universe,''
  Nucl.\ Phys.\  B {\bf 316} (1989) 391.

 A.~Tseytlin and C.~Vafa,
  ``Elements of String Cosmology,''
  Nucl.\ Phys.\  B {\bf 372} (1992) 443
  [arXiv:hep-th/9109048].



\bibitem{Hagedorn}
  R.~Hagedorn,
 ``Statistical thermodynamics of strong interactions at high-energies,''
  Nuovo Cim.\ Suppl.\  {\bf 3}, 147 (1965).

  S.~Fubini and G.~Veneziano,
  ``Level structure of dual-resonance models,''
  Nuovo Cim.\  A {\bf 64}, 811 (1969).

  K.~Bardakci and S.~Mandelstam,
  ``Analytic solution of the linear-trajectory bootstrap,''
  Phys.\ Rev.\  {\bf 184}, 1640 (1969).

  K.~Huang and S.~Weinberg,
  ``Ultimate temperature and the early universe,''
  Phys.\ Rev.\ Lett.\  {\bf 25}, 895 (1970).

  B.~Sathiapalan,
  ``Vortices on the String World Sheet and Constraints on Toral
  Compactification,''
  Phys.\ Rev.\  D {\bf 35}, 3277 (1987).

  Y.~I.~Kogan,
  ``Vortices On The World Sheet And String's Critical Dynamics,''
  JETP Lett.\  {\bf 45}, 709 (1987)
  [Pisma Zh.\ Eksp.\ Teor.\ Fiz.\  {\bf 45}, 556 (1987)].

  M.~Axenides, S.~D.~Ellis and C.~Kounnas,
 ``Universal Behavior Of D-Dimensional Superstring Models,''
  Phys.\ Rev.\  D {\bf 37}, 2964 (1988).

  D.~Kutasov and N.~Seiberg,
  ``Number Of Degrees Of Freedom, Density Of States And Tachyons In String
  Theory And Cft,''
  Nucl.\ Phys.\  B {\bf 358} (1991) 600.

  D.~Israel and V.~Niarchos,
  ``Tree-level stability without spacetime fermions: Novel examples in string
  theory,''
  JHEP {\bf 0707}, 065 (2007)
  [arXiv:0705.2140 [hep-th]]



\bibitem{AtickWitten}
  J.~Atick and E.~Witten,
  ``The Hagedorn Transition and the Number of Degrees of Freedom of String
  Theory,''
  Nucl.\ Phys.\  B {\bf 310}, 291 (1988).



\bibitem{RostKounnas}
  C.~Kounnas and B.~Rostand,
 ``Coordinate Dependent Compactifications and Discrete Symmetries,''
  Nucl.\ Phys.\  B {\bf 341} (1990) 641.




\bibitem {AKADK}
  I.~Antoniadis and C.~Kounnas,
  ``Superstring phase transition at high temperature,''
  Phys.\ Lett.\  B {\bf 261} (1991) 369.


  I.~Antoniadis, J.~P.~Derendinger and C.~Kounnas,
  ``Non-perturbative supersymmetry breaking and finite temperature
  instabilities in  $N$ = 4 superstrings,''
  arXiv:hep-th/9908137.

  C.~Kounnas,
 ``Universal thermal instabilities and the high-temperature phase of the  $N$ =
  4 superstrings,''
  arXiv:hep-th/9902072.






\bibitem{MSDS}
  C.~Kounnas,
 ``Massive Boson-Fermion Degeneracy and the Early Structure of the Universe,''
  Fortsch.\ Phys.\  {\bf 56} (2008) 1143
  [arXiv:0808.1340 [hep-th]].


 \bibitem{GravFluxes}
 C.~Angelantonj, C.~Kounnas, H.~Partouche and N.~Toumbas,
  ``Resolution of Hagedorn singularity in superstrings with gravito-magnetic
  fluxes,''
  Nucl.\ Phys.\  B {\bf 809} (2009) 291
  [arXiv:0808.1357 [hep-th]].


\bibitem{Cosmo-RT-Shifted}
  C.~Kounnas, N.~Toumbas and J.~Troost,
 ``A Wave-function for Stringy Universes,''
  JHEP {\bf 0708} (2007) 018
  arXiv:0704.1996 [hep-th].

  T.~Catelin-Jullien, C.~Kounnas, H.~Partouche and N.~Toumbas,
  ``Thermal/quantum effects and induced superstring cosmologies,''
  Nucl.\ Phys.\  B {\bf 797} (2008) 137
  arXiv:0710.3895 [hep-th].

  T.~Catelin-Jullien, C.~Kounnas, H.~Partouche and N.~Toumbas,
  ``Thermal and quantum superstring cosmologies,''
  Fortsch.\ Phys.\  {\bf 56} (2008) 792
  [arXiv:0803.2674 [hep-th]].

  T.~Catelin-Jullien, C.~Kounnas, H.~Partouche and N.~Toumbas,
 ``Induced superstring cosmologies and moduli stabilization,''
  arXiv:0901.0259 [hep-th].



\bibitem{orbifolds}
  L.~J.~Dixon, J.~A.~Harvey, C.~Vafa and E.~Witten,
  ``Strings on Orbifolds,''
  Nucl.\ Phys.\  B {\bf 261}, 678 (1985).

  L.~J.~Dixon, J.~A.~Harvey, C.~Vafa and E.~Witten,
  ``Strings on Orbifolds. 2,''
  Nucl.\ Phys.\  B {\bf 274}, 285 (1986).




\bibitem{ABK}
  I.~Antoniadis, C.~P.~Bachas and C.~Kounnas,
  ``Four-Dimensional Superstrings,''
  Nucl.\ Phys.\  B {\bf 289} (1987) 87.



\bibitem{KLT}
  H.~Kawai, D.~C.~Lewellen and S.~H.~H.~Tye,
 ``Construction of Fermionic String Models in Four-Dimensions,''
  Nucl.\ Phys.\  B {\bf 288}, 1 (1987).


\bibitem{LLS}
  W.~Lerche, D.~L\"ust and A.~N.~Schellekens,
  ``Chiral Four-Dimensional Heterotic Strings from Selfdual Lattices,''
  Nucl.\ Phys.\  B {\bf 287} (1987) 477.



\bibitem{GEPN}
  D.~Gepner and Z.~A.~Qiu,
  ``Modular Invariant Partition Functions for Parafermionic Field Theories,''
  Nucl.\ Phys.\  B {\bf 285}, 423 (1987).


  D.~Gepner,
  ``On the Spectrum of 2D Conformal Field Theories,''
  Nucl.\ Phys.\  B {\bf 287}, 111 (1987).

  D.~Gepner,
  ``Space-Time Supersymmetry in Compactified String Theory
 and Superconformal Models,''
  Nucl.\ Phys.\  B {\bf 296}, 757 (1988).

  D.~Gepner,
  ``On the algebraic structure of N=2 string theory,''
  Commun.\ Math.\ Phys.\  {\bf 142}, 433 (1991).






\bibitem{ASorbifolds}
  K.~S.~Narain, M.~H.~Sarmadi and C.~Vafa,
  ``Asymmetric Orbifolds,''
  Nucl.\ Phys.\  B {\bf 288}, 551 (1987).

  K.~S.~Narain, M.~H.~Sarmadi and C.~Vafa,
  ``Asymmetric orbifolds: Path integral and operator formulations,''
  Nucl.\ Phys.\  B {\bf 356}, 163 (1991).



\bibitem{CYtorsion}
M.~Grana, T.~W.~Grimm, H.~Jockers and J.~Louis,
  ``Soft supersymmetry breaking in Calabi-Yau orientifolds with D-branes  and
  fluxes,''
  Nucl.\ Phys.\  B {\bf 690} (2004) 21
  [arXiv:hep-th/0312232].

 D.~Lust, S.~Reffert and S.~Stieberger,
  ``Flux-induced soft supersymmetry breaking in chiral type IIb  orientifolds
  with D3/D7-branes,''
  Nucl.\ Phys.\  B {\bf 706} (2005) 3
  [arXiv:hep-th/0406092].



\bibitem{Orientifolds}
  C.~Angelantonj and A.~Sagnotti,
  ``Open strings,''
  Phys.\ Rept.\  {\bf 371} (2002) 1
  [Erratum-ibid.\  {\bf 376} (2003) 339]
  [arXiv:hep-th/0204089].


\bibitem{GeoFluxes}
J.~P.~Derendinger, C.~Kounnas, P.~M.~Petropoulos and F.~Zwirner,
  ``Superpotentials in IIA compactifications with general fluxes,''
  Nucl.\ Phys.\  B {\bf 715} (2005) 211
  [arXiv:hep-th/0411276].


  J.~P.~Derendinger, C.~Kounnas, P.~M.~Petropoulos and F.~Zwirner,
  ``Fluxes and gaugings: $N$ = 1 effective superpotentials,''
  Fortsch.\ Phys.\  {\bf 53} (2005) 926
  [arXiv:hep-th/0503229].


  G.~Villadoro and F.~Zwirner,
  ``N = 1 effective potential from dual type-IIA D6/O6 orientifolds with
  general fluxes,''
  JHEP {\bf 0506}, 047 (2005)
  [arXiv:hep-th/0503169].

  G.~Villadoro and F.~Zwirner,
  ``D terms from D-branes, gauge invariance and moduli stabilization in  flux
  compactifications,''
  JHEP {\bf 0603}, 087 (2006)
  [arXiv:hep-th/0602120].

L.~Andrianopoli, M.~A.~Lledo and M.~Trigiante,
  ``The Scherk-Schwarz mechanism as a flux compactification with internal
  torsion,''
  JHEP {\bf 0505} (2005) 051
  [arXiv:hep-th/0502083].

G.~Dall'Agata and N.~Prezas,
  ``Scherk-Schwarz reduction of M-theory on $G_2$-manifolds with fluxes,''
  JHEP {\bf 0510} (2005) 103
  [arXiv:hep-th/0509052].

  J.~P.~Derendinger, P.~M.~Petropoulos and N.~Prezas,
  ``Axionic symmetry gaugings in N = 4 supergravities and their
  higher-dimensional origin,''
  Nucl.\ Phys.\  B {\bf 785}, 115 (2007)
  [arXiv:0705.0008 [hep-th]].



 \bibitem{OpenFluxes}
  C.~Angelantonj, S.~Ferrara and M.~Trigiante,
  ``New D = 4 gauged supergravities from N = 4 orientifolds with fluxes,''
  JHEP {\bf 0310}, 015 (2003)
  [arXiv:hep-th/0306185].

  C.~Angelantonj, R.~D'Auria, S.~Ferrara and M.~Trigiante,
  ``K3 x T**2/Z(2) orientifolds with fluxes, open string moduli and  critical
 points,''
  Phys.\ Lett.\  B {\bf 583}, 331 (2004)
  [arXiv:hep-th/0312019].

  C.~Angelantonj, S.~Ferrara and M.~Trigiante,
  ``Unusual gauged supergravities from type IIA and type IIB orientifolds,''
  Phys.\ Lett.\  B {\bf 582}, 263 (2004)
  [arXiv:hep-th/0310136].

 C.~Angelantonj, R.~D'Auria, S.~Ferrara and M.~Trigiante,
  ``$K3 \times T^2/Z_2$ orientifolds with fluxes, open string moduli and  critical
points,''
  Phys.\ Lett.\  B {\bf 583} (2004) 331
  [arXiv:hep-th/0312019].

  C.~Angelantonj, M.~Cardella and N.~Irges,
  ``An alternative for moduli stabilisation,''
  Phys.\ Lett.\  B {\bf 641} (2006) 474
  [arXiv:hep-th/0608022].



\bibitem{Fluxes}
  K.~Dasgupta, G.~Rajesh and S.~Sethi,
  ``M theory, orientifolds and G-flux,''
  JHEP {\bf 9908} (1999) 023
  [arXiv:hep-th/9908088].

  A.~R.~Frey and J.~Polchinski,
  ``N = 3 warped compactifications,''
  Phys.\ Rev.\  D {\bf 65} (2002) 126009
  [arXiv:hep-th/0201029].

  S.~Gukov, C.~Vafa and E.~Witten,
  ``CFT's from Calabi-Yau four-folds,''
  Nucl.\ Phys.\  B {\bf 584} (2000) 69
  [Erratum-ibid.\  B {\bf 608} (2001) 477]
  [arXiv:hep-th/9906070].

  S.~B.~Giddings, S.~Kachru and J.~Polchinski,
  ``Hierarchies from fluxes in string compactifications,''
  Phys.\ Rev.\  D {\bf 66}, 106006 (2002)
  [arXiv:hep-th/0105097].


\bibitem{ABKW}
  I.~Antoniadis, C.~Bachas, C.~Kounnas and P.~Windey,
  ``Supersymmetry among free fermions and superstrings,''
  Phys.\ Lett.\  B {\bf 171}, 51 (1986).



\bibitem{theta12}
  C.~Kounnas, B.~Rostand and E.~T.~Tomboulis,
  ``Heterotic (2,1) supergravity in two-dimensions,''
  Nucl.\ Phys.\  B {\bf 359} (1991) 673.



\bibitem{FrShen}
  D.~Friedan, S.~H.~Shenker and E.~J.~Martinec,
  ``Covariant Quantization Of Superstrings,''
  Phys.\ Lett.\  B {\bf 160}, 55 (1985).

  D.~Friedan, E.~J.~Martinec and S.~H.~Shenker,
  ``Conformal Invariance, Supersymmetry And String Theory,''
  Nucl.\ Phys.\  B {\bf 271}, 93 (1986).

  J.~Cohn, D.~Friedan, Z.~A.~Qiu and S.~H.~Shenker,
  ``Covariant Quantization of Supersymmetric string theories"
  Nucl.\ Phys.\  B {\bf 278}, 577 (1986).


\bibitem{FKR}
  A.~E.~Faraggi, C.~Kounnas and J.~Rizos,
 ``Chiral family classification of fermionic Z(2) x Z(2) heterotic orbifold
 models,''
  Phys.\ Lett.\  B {\bf 648} (2007) 84
  [arXiv:hep-th/0606144].

  A.~E.~Faraggi, C.~Kounnas and J.~Rizos,
 ``Spinor - vector duality in fermionic Z(2) x Z(2) heterotic orbifold
 models,''
  Nucl.\ Phys.\  B {\bf 774} (2007) 208
  [arXiv:hep-th/0611251].

  A.~E.~Faraggi, C.~Kounnas and J.~Rizos,
  ``Spinor-Vector Duality in N=2 Heterotic String Vacua,''
  Nucl.\ Phys.\  B {\bf 799} (2008) 19
  [arXiv:0712.0747 [hep-th]].

  T.~Catelin-Jullien, A.~E.~Faraggi, C.~Kounnas and J.~Rizos,
  arXiv:0807.4084 [hep-th].



\bibitem{Noscale}
  E.~Cremmer, S.~Ferrara, C.~Kounnas and D.~V.~Nanopoulos,
  ``Naturally Vanishing Cosmological Constant In N=1 Supergravity,''
  Phys.\ Lett.\  B {\bf 133} (1983) 61.

  J.~R.~Ellis, C.~Kounnas and D.~V.~Nanopoulos,
  ``No Scale Supersymmetric Guts,''
  Nucl.\ Phys.\  B {\bf 247} (1984) 373.

  J.~R.~Ellis, C.~Kounnas and D.~V.~Nanopoulos,
  ``Phenomenological SU(1,1) Supergravity,''
  Nucl.\ Phys.\  B {\bf 241} (1984) 406.

  J.~R.~Ellis, A.~B.~Lahanas, D.~V.~Nanopoulos and K.~Tamvakis,
 ``No-Scale Supersymmetric Standard Model,''
  Phys.\ Lett.\  B {\bf 134}, 429 (1984).



\bibitem{StringyNoscale}
  E.~Witten,
  ``Dimensional reduction of superstring models,''
  Phys.\ Lett.\  B {\bf 155} (1985) 151.

  S.~Ferrara, C.~Kounnas and M.~Porrati,
  ``General dimensional reduction of ten-dimensional supergravity and
  superstring,''
  Phys.\ Lett.\  B {\bf 181} (1986) 263.

 M.~Cvetic, J.~Louis and B.~A.~Ovrut,
  ``A string calculation of the K\"ahler potentials for moduli of $\Z_N$
  orbifolds,''
  Phys.\ Lett.\  B {\bf 206} (1988) 227.

  L.~J.~Dixon, V.~Kaplunovsky and J.~Louis,
  ``On effective field theories describing $(2,2)$ vacua of the heterotic
  string,''
  Nucl.\ Phys.\  B {\bf 329} (1990) 27.

 M.~Cvetic, J.~Molera and B.~A.~Ovrut,
  ``K\"ahler potentials for matter scalars and moduli of $\Z_N$ orbifolds,''
  Phys.\ Rev.\  D {\bf 40} (1989) 1140.


\bibitem{ShifftedLat}
  C.~Kounnas,
  ``BPS states in superstrings with spontaneously broken SUSY,''
  Nucl.\ Phys.\ Proc.\ Suppl.\  {\bf 58} (1997) 57
  [arXiv:hep-th/9703198].


  E.~Kiritsis and C.~Kounnas,
  ``Perturbative and non-perturbative partial supersymmetry breaking: $ N$ = 4
$  \rightarrow N $= 2 $\rightarrow N$ = 1,''
  Nucl.\ Phys.\  B {\bf 503} (1997) 117
  [arXiv:hep-th/9703059].


  E.~Kiritsis, C.~Kounnas, P.~M.~Petropoulos and J.~Rizos,
  ``String threshold corrections in models with spontaneously broken
  supersymmetry,''
  Nucl.\ Phys.\  B {\bf 540} (1999) 87
  [arXiv:hep-th/9807067].


\bibitem{R-Shifted}
  J.~Scherk and J.~H.~Schwarz,
  ``Spontaneous breaking of supersymmetry through dimensional reduction,''
  Phys.\ Lett.\  B {\bf 82} (1979) 60.

  R.~Rohm,
  ``Spontaneous supersymmetry breaking in supersymmetric string theories,''
  Nucl.\ Phys.\  B {\bf 237} (1984) 553.

  C.~Kounnas and M.~Porrati,
  ``Spontaneous Supersymmetry Breaking in String Theory,''
  Nucl.\ Phys.\  B {\bf 310} (1988) 355.

  S.~Ferrara, C.~Kounnas and M.~Porrati,
  ``N=1 Superstrings With Spontaneously Broken Symmetries,''
  Phys.\ Lett.\  B {\bf 206} (1988) 25.

  S.~Ferrara, C.~Kounnas, M.~Porrati and F.~Zwirner,
  ``Effective Superhiggs and Strm**2 from Four-dimensional Strings,''
  Phys.\ Lett.\  B {\bf 194} (1987) 366.

  S.~Ferrara, C.~Kounnas and M.~Porrati,
``Superstring Solutions with Spontaneously Broken Four-Dimensional
  Supersymmetry,''
  Nucl.\ Phys.\  B {\bf 304} (1988) 500.



\bibitem{WittenHoloFactor}
  E.~Witten,
 ``Three-Dimensional Gravity Revisited,''
  arXiv:0706.3359 [hep-th].



\bibitem{HennStrominger}
  J.~D.~Brown and M.~Henneaux,
  ``Central Charges in the Canonical Realization of Asymptotic Symmetries: An
  Example from Three-Dimensional Gravity,''
  Commun.\ Math.\ Phys.\  {\bf 104}, 207 (1986).

  D.~Anninos, W.~Li, M.~Padi, W.~Song and A.~Strominger,
  ``Warped $AdS_3$ Black Holes,''
  arXiv:0807.3040 [hep-th].


\end{thebibliography}
\end{document}